\documentstyle[epsf,psfig,aps]{revtex}
\newcommand{\bea}{\begin{eqnarray}}
\newcommand{\eea}{\end{eqnarray}}
\newcommand{\be}{\begin{equation}}
\newcommand{\ee}{\end{equation}}
\newcommand{\bt}{\begin{tabular}}
\newcommand{\et}{\end{tabular}}
\newcommand{\Tr}{{\rm Tr}}
\newcommand{\no}{\nonumber}
\newcommand{\ovl}{\overline}

\newcommand{\pa}{\partial}
\newcommand{\beas}{\begin{eqnarray*}}
\newcommand{\eeas}{\end{eqnarray*}}

\begin{document}
\title{Nuclei in a chiral $SU(3)$ model
}
\author{P. Papazoglou$^1$, D. Zschiesche$^1$, S. Schramm$^2$, J. Schaffner-Bielich$^3$, 
        H. St\"ocker$^1$, and W. Greiner$^1$}
  \address{
        $^1$Institut f\"ur Theoretische Physik, 
        $^{~}$Postfach 11 19 32, D-60054 Frankfurt am Main, Germany\\
        $^2$GSI Darmstadt, Postfach 11 05 52, D-64220 Darmstadt, Germany\\
        $^3$Nuclear Science Division, LBNL, 1 Cyclotron Road, Berkeley, 
California 94720
} 
\date{\today}
\maketitle
\begin{abstract}
 Nuclei can be described satisfactorily in a nonlinear chiral $SU(3)$-framework, 
  even with standard potentials of the linear $\sigma$-model.
The condensate value of the strange scalar meson is found to be 
important for the properties of nuclei even without adding hyperons.
By neglecting terms which couple the strange to the nonstrange condensate 
 one can reduce the model
to a Walecka model structure embedded in $SU(3)$.
We discuss inherent problems with chiral $SU(3)$ models regarding
hyperon optical potentials.
\end{abstract}
\draft
\section{Introduction}
Recently, the general principles of chiral symmetry and broken 
scale invariance in QCD have received renewed attention at finite 
baryon densities.
There the underlying theory of strong interactions, QCD, is however
not solvable in the nonperturbative  low energy regime. However, 
QCD constraints can be imposed on an effective ansatz for nuclear theory 
through symmetries that determine largely how the hadrons should interact 
with each other. In this spirit, models with  
SU(2)$_L \times $SU(2)$_R$ symmetry and scale invariance were 
applied to nuclear matter at zero and finite temperature 
and to finite nuclei \cite{heid94,cart95,furn95,mish93,paper1}. 
As a new feature, a glueball field $\chi$, the dilaton, was included 
which accounted  for the broken 
scale invariance of QCD at tree level through a logarithmic potential \cite{sche80}.
The success of these models  established 
the applicability of this approach  to the  relativistic description of 
the nuclear many-body problem.  

Chiral SU(3) models have been quite successful in modelling hadron
interactions. Meson-meson interactions can be described satisfactorily
by using the linear SU(3) $\sigma$ model \cite{Torn97}. 
Kaon-nucleon scattering data can be well reproduced using
a chiral effective SU(3) Lagrangian \cite{Waas97} using a Lippmann-Schwinger
approach \cite{Koch94}. The lowest order term is sufficient to describe the
kaon-nucleon scattering data when including relativistic effects consistently 
and adding the $\eta$ channel \cite{Oset97}. Especially, the in-medium
properties of the kaon in nuclear matter are of considerable interest for 
recent measurements of kaon spectra at GSI, Darmstadt at subthreshold energies
\cite{kaos}. All of the above models lack the feature of including the
nucleon-nucleon interaction on the same chiral SU(3) basis and therefore do not
provide a consistent extrapolation to finite density. 

Within
SU(3) chiral models one can also take a different view at the properties of
metastable exotic multihypernuclear objects \cite{Scha92}. 
The relativistic
mean field model was extended to include the baryon octet and the vector
nonet by using SU(6)-symmetry for the coupling constants. The existence of
strange hadronic matter and 
bound objects consisting purely of hyperons has been proposed
\cite{Scha93,Scha94}. The properties of strange hadronic matter are remarkably
close to those of strangelets and can be negatively charged while carrying a
positive baryon number. This has certain impacts for present heavy-ion searches
looking for strangelets and other possible exotica \cite{e864,na52}.
  
We have recently extended the chiral effective model to 
SU(3)$_L\times$SU(3)$_R$ \cite{paper2} including the baryon octet.  
This approach shall provide a basis to shed light on the properties of strange
hadrons, as the in-medium properties of the kaon and the properties of
strange hadronic matter, by pinning down the nuclear force in a chiral
invariant way. 
This paper continues our previous work \cite{paper2}, 
which has applied a linear realization of chiral SU(3) symmetry and the 
concept of broken scale invariance to the 
description of hadronic matter in the vacuum and in the medium. 
It has been found that simultaneously both 
hadronic masses of the various SU(3) multiplets 
and the nuclear matter equation of state can be described reasonably well
within a model respecting chiral symmetry.
 
However, it has been shown that the central potentials of the hyperons come 
out too large. They  could not be
corrected within a model with Yukawa-type baryon-meson interactions.
The reason for this deficiency is threefold \cite{paper2}:
\begin{itemize}
\item Firstly, linear realizations of chiral symmetry restrict
the coupling of the spin-0 mesons to the baryons to be symmetric (d-type), 
while the spin-1 mesons are coupled to baryons antisymmetrically (f-type).
This destroys the balance between the repulsive contribution of the vector potential 
and the attraction due to the scalar potential. Therefore,  
the hyperon potentials attain too large values.
\item Secondly, the  condensate of the strange scalar meson $\zeta$ 
changes considerably in the nuclear medium 
even for zero strangeness within this approach, 
because it couples to the non-strange scalar field $\sigma$ 
and therefore provides additional attraction. This is not 
counterbalanced by repulsive contributions from 
the strange vector field $\phi_{\mu}$, since $\phi_\mu$ does neither couple 
to the nucleon nor to the $\omega_{\mu}$-field.
\item Thirdly, it is not possible to correct these values of the 
hyperon potentials through 
explicit symmetry-breaking terms, because they would destroy the relations for the partially 
conserved axial-vector currents (PCAC) of the pion and the kaon. 
\end{itemize}

In order to deal with this general problem, nonlinear interaction terms 
of baryons with mesons were introduced
in \cite{paper2} in a chirally invariant way. However, although 
a cubic interaction of baryons with spin-0 mesons (with strong 
coupling of the strange condensate to the nucleons) produces 
reasonable hyperon potentials, such a  form for this interaction 
seems quite artificial.
Furthermore, the high mass of the strange meson ($\approx 1$ GeV)  
excludes a reasonable description of nuclei, since the oscillations in the 
charge density are too high for such an ansatz. 
Hence, although the cubic fit works satisfactorily for nuclear matter, it is 
not suitable to describe finite nuclei.
The constraints imposed by the linear realization of chiral 
symmetry do not allow for a  simultaneous description of both, finite nuclei 
and hyperon potentials.

In this paper, we propose the adoption of the {\it nonlinear realization of chiral symmetry}
as a solution to this problem. 
As was proven in \cite{ccwz}, it 
is sufficient to have a local SU(3) invariance for the hadrons, 
with the pseudoscalar mesons appearing only in derivative couplings. 
Therefore, both the d-type and the f-type of coupling is possible 
between baryons and 
scalar mesons. In addition, the pseudoscalar meson 
masses then depend only on the explicit symmetry breaking term. 
There is no reference to pseudoscalar mesons in the chirally invariant 
potential. Therefore, the potential only determines the  
masses of the scalar mesons. 
This allows to decouple  the strange 
condensate from the nonstrange one. Then both the results 
obtained with the SU(2) chirally symmetry models \cite{heid94}
and those of the nonlinear $\sigma-\omega$ model \cite{boguta77,fpw} 
can be reproduced as special realizations of the present general 
chiral SU(3) model. One can then systematically add terms of
strange-to nonstrange condensate coupling. Therefore, one can study the  
limiting case of a system consisting of nucleons only. Furthermore, 
explicit symmetry breaking terms (e.g. to correct the hyperon 
potentials) can be added without altering the PCAC-relations of the pion 
and the kaon.

In this work it is demonstrated
that one can simultaneously describe nuclei and the properties 
of strange hadrons within the framework of the nonlinear realization of 
chiral symmetry.
Formally the sectors of scalar and pseudoscalar mesons are decoupled. 
However, the analysis \cite{paper2} and the closeness of the coupling 
constant  $g_{N \sigma}$ to $m_N/f_{\pi}$ in the Boguta-Bodmer model
seem to suggest to keep the  
constraints of the decay constants of the kaons and pions imposed 
on the vacuum expectation values (VEV) of the nonstrange and strange scalar
fields 
$\sigma$ and $\zeta$.

Our paper is structured as follows.
The nonlinear realization and the connection between the linear and the nonlinear 
$\sigma$ model
of chiral symmetry are introduced in Sec. \ref{theory}.
The chiral SU(3) Lagrangian is constructed and discussed in 
Sec. \ref{lagrangian}. The equations of motion are solved in the 
mean field approximation which is described in Sec. \ref{mfa}.
In Sec. \ref{work} various parameter sets are presented which all account 
for a satisfactory description of finite nuclei:
These include a Lagrangian with the potential of the linear 
SU(3) $\sigma$ model as constructed 
in \cite{paper2} with a  modified 
baryon scalar-meson interaction. In
the limit that the strange and nonstrange condensates are decoupled,
the SU(2) chiral models \cite{heid94} and \cite{fpw,tang95} are recovered, 
but embedded in the nonlinearly realized chiral SU(3) framework.

\section{The nonlinear realization of chiral symmetry}
\label{theory}
In some neighborhood of the identity transformation,  
every group element $g'(x)$ of a compact, 
semi-simple group $G$ with a subgroup $H$ can be decomposed
uniquely into a product of the form \cite{ccwz}
\be
\label{cgruppe}
g'(x) =\exp\left[ i \sum \xi_a(x) S_a \right] \exp \left[ i \sum \theta_b(x) T_b \right]
\equiv u\left(\xi_a(x)\right) h\left(\theta_b(x)\right),
\ee
where $h(\theta_b)$ is an element of $H$. $\xi_a$ and $\theta_b$ are parameters of the 
symmetry transformation which are generally space-time dependent. 
$S_a$ and $T_b$ represent the generators of the group G.

For the case of SU(3)$_L\times$ SU(3)$_R$ symmetry, the generators 
are the vectorial ($T_b=Q_b$) and axial ($S_a=Q_a^5$) charges, 
respectively, and the subgroup is $H=SU(3)_V$.

For our model, we assume invariance under {\it global} $SU(3)_L\times SU(3)_R$ 
transformations, 
\be
 g =\exp\left[ i \sum \alpha_L^a \lambda_{L a} \right] \exp \left[ i \sum 
\alpha_R^b  \lambda_{R b} \right]
\equiv L\left(\alpha_L\right) R\left(\alpha_R\right).
\ee
Here, the representation of Gell-Mann matrices 
$\lambda_L=\lambda (1-\gamma_5)/2$ and 
$\lambda_R=\lambda (1+\gamma_5)/2$ with space-time independent
parameters $\alpha_L$ and $\alpha_R$ is used.

The product $g \, u(\xi_a(x))$ is still
an element of $G$ and can be written as
\be
\label{gu}
g\, \exp \left[ i \sum \xi_a S_a \right]=
\exp \left[ i \sum \xi_a^{\prime}(g,\xi_a) S_a \right] 
\exp \left[ i \sum \theta_b^{\prime}(g,\xi_a) T_b \right] , 
\ee
where, in general, both $\xi_a^{\prime}$ and $\theta_b^{\prime}$ depend on $g$ and $\xi_a$.
Let
\be
\tilde q \rightarrow D(h) \tilde q
\ee
be a linear representation of the subgroup $H$ of $G$. Then the 
transformation
\be
\label{cole6}
g: \xi \rightarrow \xi^{\prime}, 
\tilde{q} \rightarrow D\left(\exp\left[ i \sum \theta_b^{\prime} T_b \right]\right)
\tilde{q}
\ee
constitutes a nonlinear realization of $G$.

 The local parameters of the axial charges are 
identified with the fields of the pseudoscalar mesons \cite{weinbuch}. 
In the representation of Gell-Mann matrices one has 
(see also appendix \ref{append})
\be
\label{piexp}
u(\pi_a(x)) = \exp\left[\frac{i}{2 \sigma_0} \pi^a(x) \lambda_a \gamma_5\right] .
\ee
This assignment has the advantage 
that the pseudoscalar mesons are the parameters of the symmetry transformation. 
They will therefore only appear if the symmetry is explicitly broken or 
in terms with derivatives of the fields.

The composition of hadrons in terms of its constituents, the quarks, 
has to be determined  
in order to build models with hadronic degrees of freedom. 
This strategy has been followed e.g. in \cite{paper2} and is 
adopted also here. The transformation properties of the hadrons 
in the nonlinear representation can be derived if the `old' quarks $q$
are related to the `new' quarks $\tilde q$ of the nonlinear 
representation.

The quarks of the nonlinear representation transform with the 
vectorial subgroup $SU(3)_V$ in 
accord with equation (\ref{cgruppe}). Splitting the quarks in left- and 
right-handed parts, they can be written as
\be
q_L=u \tilde{q}_L \qquad  q_R=u^\dagger \tilde{q}_R  .
\ee 
These equations are connected by parity.
The ambiguity in the choice of $h$ is avoided by setting $h=1$.
The transformation properties of the pions and the new quarks are found 
by considering how the old quarks 
transform:
\bea
\label{qtrafo}
q' =L q_L + R q_R = L u \tilde{q}_L+ R u^\dagger \tilde{q}_R .
\eea
According to (\ref{gu}), (set $g=L$), 
\be
\label{uprime}
     Lu=u'h  \quad ; \quad  R u^\dagger=u{^\dagger}'  h,         
\ee
where the right equation is the parity transformed one of the left equation.
Here and in the following, the abbreviations
$u\equiv u(\pi_a(x))$ and $u'\equiv u(\pi_a'(x))$ are used.
By inserting these relations into (\ref{qtrafo}), one sees that $\tilde{q}$ 
transforms with $SU(3)_V$ as 
\be
  \tilde{q}_L'= h\tilde{q}_L \quad ; \quad  
  \tilde{q}_R'= h \tilde{q}_R. 
\ee
According to (\ref{gu}), in general the vector transformation is  a local, nonlinear function 
depending on pseudoscalar mesons, $h=h(g,\pi_a(x))$.
Following equation (\ref{uprime}), the pseudoscalar mesons transform nonlinearly as
\bea
\label{utrans}
u'          &=& L u h^{\dagger}= h u R^{\dagger}\\
u^{\dagger} &=& h u^{\dagger}L^{\dagger} = R u^{\dagger} h^{\dagger}. 
\eea
The second set of equalities are again due to parity.
In contrast to the linear realization of  chiral symmetry, there 
is no distinction between 
the left and right space. Therefore, only the representations 
{\bf 8} and {\bf 1}
of the lowest-lying hadrons are possible. The various octets transform 
accordingly, e.g. for the scalar ($X$), 
vector ($V_\mu=l_{\mu}+r_\mu$), axial vector (${\cal A}_\mu=l_\mu-r_\mu$)
and baryon ($B$) matrices one has,
\be
     X'       =h X       h^\dagger   \qquad 
     V_{\mu}' =h V_{\mu} h^\dagger   \qquad 
     {\cal A}_{\mu}' =h {\cal A}_\mu   h^\dagger   \qquad 
     B'       =h B       h^\dagger .
\ee
The present, nonlinearly transforming, hadronic fields can be obtained from the linearly transforming 
ones described in \cite{paper2} by multiplying them by $u(\pi(x))$ and its conjugate:
(see also \cite{stoks})
\bea
\label{fields}
  X &=& \frac{1}{2}(u^\dagger M u^\dagger +u M^\dagger u)  \qquad 
  Y =\frac{1}{2}(u^\dagger M u^\dagger -u M^\dagger u)   \\ 
  l_{\mu} &=& u^\dagger \tilde l_{\mu} u \qquad r_\mu =u\tilde r_\mu u^\dagger \\
  B_L &=& u^\dagger \Psi_L u \qquad B_R = u \Psi_R u^\dagger    .
\eea              
Here, $M=\Sigma+i\Pi$ and its conjugate contains the nonets of the linearly transforming 
scalar ($\Sigma$) and pseudoscalar ($\Pi$) mesons, whereas $\tilde l_\mu$, $\tilde l_\mu$, $\Psi_L$, and 
$\Psi_R$ are the left and right-handed parts of the spin-1 mesons and baryons in the 
linear representation, respectively. 

\section{Lagrangian}
\label{lagrangian}
In this section, the various terms of the Lagrangian
\be
\label{lagrange}
{\cal L} = {\cal L}_{\mathrm{kin}}+\sum_{W=X,Y,V,{\cal A},u}{\cal L}_{\mathrm{BW}}+
{\cal L}_{\mathrm{VP}}
+{\cal L}_{\mathrm{vec}}+{\cal L}_{0}+{\cal L}_{\mathrm{SB}} \no
\ee
are discussed in detail. ${\cal L}_{\mathrm{kin}}$ is 
the kinetic energy term, ${\cal L}_{\mathrm{BW}}$ includes the  
interaction terms of the different baryons with the various spin-0 and spin-1 
mesons and with the photons. In ${\cal L}_{\rm{VP}}$, the interaction terms 
of vector mesons with pseudoscalar mesons and with photons is summarized. 
 ${\cal L}_{\rm{vec}}$ generates the 
masses of the spin-1 mesons through 
interactions with spin-0 mesons, and ${\cal L}_{0}$ gives the meson-meson 
interaction terms which induce the spontaneous breaking of chiral symmetry.
It also includes the scale breaking logarithmic potential. Finally, 
${\cal L}_{\mathrm{SB}}$ introduces an explicit symmetry breaking of the
U(1)$_A$, the SU(3)$_V$, and the chiral symmetry. 
\subsection{Kinetic energy terms}
Since the vector transformation $h(\pi(x))$ of the hadrons depends in general 
on the pseudoscalar mesons and thus is local, 
covariant derivatives have to 
be used for the kinetic terms  in order to preserve chiral invariance.
The covariant derivative, i.e., for the baryons,  reads
\be
\label{kovder}
D_{\mu}B = \partial_{\mu}B +i [\Gamma_{\mu}, B]   
\ee
where 
\be
\label{gammamu}
\Gamma_{\mu} =-\frac{i}{2} \left[u^{\dagger}(\partial_{\mu}+ig_v l_{\mu}) u
+u (\partial_{\mu}+ig_v r_{\mu}) u^\dagger \right]. 
\ee
This is a composite vector-type field, which transforms according to 
\bea
\Gamma_{\mu}'&=& h\Gamma_{\mu}h^{\dagger}-ih \pa_{\mu} h^{\dagger} 
\eea
The external fields $l_{\mu}$ and $r_{\mu}$ can be any gauge field 
of the weak or the electromagnetic interactions. The spin-1 nonet 
of the strong interactions are here introduced as massive, homogeneously transforming 
fields, following 
the approach \cite{borasoy,birse}, in order to avoid complications arising 
from the mixing of the axial with the pseudoscalar mesons. 
Therefore, only the photon gauge field,  
$l_{\mu}=r_{\mu}= Q A_{\mu}$, with $Q=T_3+Y/2$, and the electrical charge,
$g_v=e$, will be incorporated into the present approach.

The kinetic energy term of the pseudoscalar mesons is introduced by 
defining (in analogy to Eq. (\ref{gammamu})) the axial-vector by 
\be
u_{\mu} =-\frac{i}{2} \left[u^{\dagger}(\partial_{\mu}+ig_v l_{\mu}) u
-u (\partial_{\mu}+ig_v r_{\mu}) u^\dagger \right],  
\ee
which transforms as $u_{\mu}'   = h u_{\mu} h^{\dagger}$.

The standard form for the kinetic energy of the pseudoscalar mesons is 
$\Tr (u_\mu u^\mu)$. However, the approximate validity 
of $g_{N \sigma}\approx m_N/f_\pi$, where $f_\pi=-\sigma_0$, in 
the Walecka-type models \cite{fpw,tang95} and results obtained in \cite{paper2,heid94} 
indicate that the constraints of the linear $\sigma$ model on the scalar condensates 
in the vacuum, are also applicable to the description of hadronic matter and nuclei.
To incorporate those constraints in the nonlinear realization we modify 
the standard kinetic energy term to include a coupling of the scalar and pseudoscalar 
mesons, 
\be
\label{pikin}
  \Tr (u_{\mu} X u^{\mu}X +X u_{\mu} u^{\mu} X) . 
\ee
This term contains, besides higher order self-interactions,
the kinetic energy term for the pseudoscalar mesons if the pseudoscalar 
matrix in the exponential of Eq. \ref{piexp} is defined as
\be
\label{psmatrix}
\frac{1}{\sqrt{2}}\pi_a \lambda^a
=\left (\begin {array}{ccc} 
  \frac{1}{\sqrt{2}}\left ( \pi^0+{\frac {\eta^8}{\sqrt {1+2\,{w}^{2}}}}\right )&\pi^{+}
 &2\,{\frac {K^{+}}{w+1}}\\
  \noalign{\medskip}\pi^{-}&\frac{1}{\sqrt{2}}\left 
 (- \pi^0+
 {\frac {\eta^8}{\sqrt {1+2\,{w}^{2}}}}\right )&2\,{\frac { K^0
  }{w+1}}\\\noalign{\medskip}2\,{\frac {K^-}{w+1}}&2\,{\frac { 
 \ovl{K}^0}{w+1}}&-{\frac {\eta^8\,\sqrt {2}}{\sqrt {1+2\,{w}^{2}}}}
 \end {array}\right ) .
\ee
The renormalization factors containing
$w=\sqrt{2}\zeta_0/\sigma_0$ are included 
to obtain  the canonical form 
of the kinetic energy terms for pseudoscalar mesons\footnote{
 The same normalization 
of the pseudoscalar matrix has to be taken if the kinetic energy term 
$\frac{1}{2}\Tr(\partial_{\mu} M^{\dagger} \partial^{\mu} M)$ is used
with 
\be
\label{mmdagger}
M=u(X+i Y)u, \quad  M=u^\dagger(X-i Y)u^\dagger
\ee
 substituted.} from (\ref{pikin}). 
For $w=1$, one 
has an $SU(3)_V$ symmetric vacuum and the matrix (\ref{psmatrix})
reduces to the matrix normally used e.g. 
in chiral perturbation theory \cite{holstein}. The advantage of (\ref{psmatrix})
is that $SU(3)_V$ breaking effects (such as $f_{\pi}\ne f_K$) are accounted 
for even at lowest order.

After computing the axial current for pions and kaons from Eq. (\ref{pikin}), 
one obtains the same relations 
\be
\label{zeta0}
\sigma_0 = -f_{\pi} \qquad \zeta_0 = -\frac{1}{\sqrt{2}}(2 f_K - f_{\pi}) ,
\ee 
for the VEV of the scalar condensates found in the linear $\sigma$-model 
\cite{paper2}. 

In order to construct a chirally invariant kinetic term for 
the spin-1 mesons, the ordinary derivatives must be replaced 
by the covariant derivatives as defined in Eq. (\ref{kovder}), 
\be
   V_{\mu \nu} =D_{\mu} V_{\nu}-D_{\nu} V_{\mu}, 
\ee
and analogously for the axial vector mesons, where the symbol ${\cal A}_{\mu \nu}$ is used.
 
In summary, the kinetic energy terms read
\bea
\label{kinetic}
{\cal L}_{kin} &=& i\Tr \overline{B} \gamma_{\mu} D^{\mu}B 
                + \frac{1}{2} \Tr D_{\mu} X D^{\mu} X             
+  \Tr (u_{\mu} X u^{\mu}X +X u_{\mu} u^{\mu} X)
                + \frac{1}{2}\Tr D_{\mu} Y D^{\mu} Y\\
               &+&\frac {1}{2} D_{\mu} \chi D^{\mu} \chi 
                - \frac{ 1 }{ 4 } \Tr \left(V_{ \mu \nu } V^{\mu \nu }  \right)
- \frac{ 1 }{ 4 } \Tr \left(F_{ \mu \nu } F^{\mu \nu }  \right)
- \frac{ 1 }{ 4 } \Tr \left( {\cal A}_{ \mu \nu } {\cal A}^{\mu \nu }  \right)
\eea
where we included
the usual field strength tensor of the photon
$F_{\mu \nu}$ as we want to discuss electromagnetic form factors and nuclei later on. 
The pseudoscalar singlet is independent of the octet and has 
thus a kinetic term of its own. For the dilaton field $\chi$ 
(for its discussion see Sec. \ref{scale}), which 
is also a chiral singlet, it makes no difference if the  normal derivative is replaced with 
the covariant derivative  because the additional commutator term vanishes. 
\subsection{Baryon-meson interaction}
The various interaction terms of baryons  with 
mesons are discussed in this section. 
The $SU(3)$ structure of the 
the baryon-meson interaction terms are the same for all mesons, 
except for the difference in Lorentz space. 
For a general meson field $W$ they read
\be
{\cal L}_{\mbox{BW}} = 
-\sqrt{2}g_8^W \left(\alpha_W[\ovl{B}{\cal O}BW]_F+ (1-\alpha_W) 
[\ovl{B} {\cal O}B W]_D \right)
- g_1^W \frac{1}{\sqrt{3}} \Tr(\ovl{B}{\cal O} B)\Tr W  \, ,  
\ee
with $[\ovl{B}{\cal O}BW]_F:=\Tr(\ovl{B}{\cal O}WB-\ovl{B}{\cal O}BW)$ and 
$[\ovl{B}{\cal O}BW]_D:= \Tr(\ovl{B}{\cal O}WB+\ovl{B}{\cal O}BW) - \frac{2}{3} 
\Tr (\ovl{B}{\cal O} B) \Tr W$.
The different terms to be considered are those for the interaction
of baryons,  with
scalar mesons ($W=X, {\cal O}=1$), with 
vector mesons  ($W=V_{\mu}, {\cal O}=\gamma_{\mu}$ for the vector and 
$W=V_{\mu \nu}, {\cal O}=\sigma^{\mu \nu}$ for the tensor interaction),
with axial vector mesons ($W={\cal A}_\mu, {\cal O}=\gamma_\mu \gamma_5$)
and with
pseudoscalar mesons ($W=u_{\mu},{\cal O}=\gamma_{\mu}\gamma_5$), respectively.
For $u_{\mu}$, the singlet term is vanishing, because
the matrix in the exponential of (\ref{piexp}) 
involves only the pseudoscalar octet and is thus traceless. 
The interaction of the pseudoscalar chiral singlet $Y$ with baryons has the 
structure $g^Y_1 \Tr(\ovl{B}\gamma_{\mu}\gamma_5 B)\Tr Y$.

Since the pseudoscalar mesons are solely contained in the exponential, 
the only possible form of their coupling with baryons is the pseudovector 
interaction.
There, the coupling constant $g_A\equiv \sqrt{2} g_8^u$ 
is restricted by the Goldberger-Treiman relation. In contrast 
to the linear representation, the axial coupling constant $g_A$ 
is not unity, but a value $g_A\simeq 1.26$ can be assigned to it. 
In this case the mixing angle between $f$-type and $d$-type coupling 
is $\alpha_u \simeq$0.4 \cite{bourquin}.

The remaining coupling constants 
will be discussed in the following\footnote{The tensor 
coupling will not be considered further in the present calculations, 
although it may be important in particular
for the description of exotic nuclei and hypernuclei.}.

\subsubsection{Scalar mesons}
\label{bmver}
The baryons and the scalar mesons 
transform equally in the left and right subspace.  
Therefore, in contrast to the linear realization of chiral symmetry, 
a $f$-type coupling is allowed for the baryon-meson interaction. 
In addition, it is possible to construct mass terms for 
baryons and to couple them to chiral singlets. 
After insertion of the vacuum matrix 
$\langle X\rangle$, (Eq. (\ref{vev})), one obtains the baryon masses as 
generated by the VEV of the two meson fields:
\bea
\label{bmver1}
 m_N        &=& m_0 -\frac{1}{3}g_8^S(4\alpha_S-1)(\sqrt{2}\zeta-\sigma) \\ \no
 m_{\Lambda}&=& m_0-\frac{2}{3}g_8^S(\alpha_S-1)(\sqrt{2}\zeta-\sigma) \\ \no
 m_{\Sigma} &=& m_0+\frac{2}{3}g_8^S(\alpha_S-1)(\sqrt{2}\zeta-\sigma)  \\ \no
 m_{\Xi}    &=& m_0+\frac{1}{3}g_8^S(2\alpha_S+1)(\sqrt{2} \zeta-\sigma) \no
\eea
with $m_0=g_1^S(\sqrt{2} \sigma+\zeta)/\sqrt{3}$.
The three parameters $g_1^S$, $g_8^S$ and $\alpha_S$ can be used to fit the 
baryon masses 
to their experimental values. Then, besides the current quark mass terms discussed in 
Sec. \ref{esb}, no additional explicit symmetry 
breaking term is needed. Note that the nucleon mass depends on the 
{\it strange condensate $\zeta$!} For $\zeta=\sigma/\sqrt{2}$ (i.e. 
$f_{\pi}=f_K$), the masses are degenerate, and the vacuum is 
SU(3)$_V$-invariant.

It is desirable to have an alternative way of baryon mass generation, 
where the nucleon mass depends only on $\sigma$. This can be accomplished 
by taking the limit $\alpha_S=1$ and $g_1^S=\sqrt{6}g_8^S$. Then, 
the coupling constants between the baryons and the two scalar condensates 
are related to the additive quark model. This leaves only one coupling 
constant free to adjust for the correct nucleon mass. For a fine-tuning 
of the remaining masses, it is necessary to introduce an explicit 
symmetry breaking term, which breaks the SU(3)-symmetry along the hypercharge 
direction. A possible term already discussed in \cite{paper2,sche69}, which respects the 
Gell-Mann-Okubo mass relation, is
\be 
\label{sche-esb}
  {\cal L}_{\Delta m} = -m_1 \Tr (\ovl{B} B - \ovl{B} B S)
  -m_2 \Tr (\ovl{B} S B)
\ee
where $S_{b}^a = -\frac{1}{3}[\sqrt{3} (\lambda_8)_{b}^a-\delta_{b}^a]$. 
As in the first case, the three coupling constants $g_{N \sigma}\equiv 3 g_8^S$, 
$m_1$ and $m_2$ are sufficient to reproduce the experimentally known 
baryon masses. Explicitly, the baryon masses have the values
\bea
\label{bmver2}
m_N &=& -g_{N \sigma}  \sigma   \\ \no
m_{\Xi} &=&-\frac{1}{3} g_{N \sigma} \sigma -\frac{2}{3} g_{N \sigma} \sqrt{2}\zeta      +m_1  +m_2\\ \no
m_{\Lambda} &=& -\frac{2}{3} g_{N \sigma} \sigma-\frac{1}{3} g_{N \sigma} \sqrt{2}\zeta+\frac{m_1+2 m_2}{3} \\ \no
m_{\Sigma} &=& -\frac{2}{3} g_{N \sigma} \sigma -\frac{1}{3} g_{N \sigma} \sqrt{2}\zeta +m_1 , 
\eea   
For both versions of baryon-meson interaction the parameters are fixed to yield the 
baryon masses $m_N=939$ MeV, $m_\Lambda=1115$ MeV, $m_\Sigma=1196$ MeV, and 
$m_\Xi=1331.5$ MeV.

\subsubsection{Vector mesons}
Two independent interaction 
terms of baryons with spin-1 mesons can be constructed
in analogy with the baryon-spin-0-meson interaction . They correspond to 
the antisymmetric ($f$-type) and symmetric ($d$-type) couplings, 
respectively. 
From the universality 
principle \cite{saku69} and the vector meson dominance model one may 
conclude that the $d$-type 
coupling should be small. For most of the fits $\alpha_V=1$, 
i.e. $f$-type coupling, 
is used. However, a small admixture of d-type coupling 
allows for some fine-tuning of the single particle energy levels of nucleons 
in nuclei (see below). 

As for the case with scalar mesons \ref{bmver}, for $g_1^V=\sqrt{6}g_8^V$, the strange vector field $\phi_{\mu} \sim  
\ovl{s}\gamma_{\mu} s $ 
does not couple to the nucleon implying that the strange vector form 
factor of the nucleon is very small. The remaining 
couplings to the strange baryons are then determined by symmetry relations:
\bea
          g_{N N \omega} &=&(4 \alpha_V-1) g_8^V \\ \no
          g_{\Lambda \Lambda \omega} &=& \frac{2}{3} (5 \alpha_V-2)g_8^V  
\qquad  g_{\Lambda \Lambda \phi}    = - \frac{\sqrt{2}}{3} (2 \alpha_V+1)g_8^V
 \\ \no  
         g_{\Sigma \Sigma \omega} &=& 2 \alpha_V g_8^V 
  \qquad g_{\Sigma \Sigma \phi} = - \sqrt{2} (2\alpha_V-1)g_8^V \\ \no
     g_{\Xi \Xi \omega} &=& (2 \alpha_V-1) g_8^V 
\qquad  g_{\Xi \Xi \phi} = -2 \sqrt{2} \alpha_V g_8^V \quad .           
\eea
In the limit $\alpha_V=1$, the relative values of the coupling constants 
are related to the additive quark model via: 
\be
\label{quarkcoupling}
 g_{\Lambda \omega} = g_{\Sigma \omega} = 2 g_{\Xi \omega} = \frac{2}{3} 
 g_{N \omega}=2 g_8^V \qquad 
 g_{\Lambda \phi} = g_{\Sigma  \phi} = \frac{g_{\Xi \phi}}{2} = 
   \frac{\sqrt{2}}{3} g_{N \omega}  .
\ee   
Note that all coupling constants are fixed once e.g. $g_{N\omega}$ is specified.

Since the axial vector mesons  have a vanishing expectation value at the mean-field 
level their coupling constants to the baryons will not be discussed here.

\subsection{Electromagnetic structure of pseudoscalar mesons}
The interaction Lagrangian of the vector mesons 
with pions and the photon takes the form
\be
   {\cal L}_{VP}=e \Tr\left(A_\mu \Gamma^\mu \right)+g \Tr\left(V_\mu\Gamma^\mu \right)
+\frac{e}{4 g_\gamma} F_{\mu \nu}\Tr\left[ 
\left(u^\dagger Q u + u Q u^\dagger \right)  V^{\mu \nu} \right]. 
\ee 
The first term originates from the kinetic energy term (\ref{pikin}). 
The remaining two terms can be motivated from a gauge and chiral invariant
Lagrangian approach \cite{Schechter86,Klingl96}.

One obtains for the form factor of the pion ($Q^2=-q^2$):
\be
   F_\pi(Q^2)=1-\frac{g}{g_\gamma} \frac{Q^2}{Q^2+m_{\rho}^2}
\ee
Note that the pion does
not couple to the other vector mesons $\omega$ and $\phi$. 

With $g=6.05$ from $\rho^0 \rightarrow \pi^+ \pi^-$ 
\cite{pdg96} and $g_\gamma=5.04$ from $\rho^0 \rightarrow e^+ e^-$
the mean-square charge radius of the pion is
\be
   \langle r_\pi^2 \rangle \equiv -6 \frac{dF(Q^2)}{d Q^2}|_{Q^2=0} =
0.48 \rm \, fm^{2} .
\ee
Experimentally, $\langle r_\pi^2 \rangle =0.432\pm0.016 \, \rm fm^{2}$
\cite{schlumpf}. 

In analogy to the pion, one obtains for the form factor of the kaon 
\be
F_K{^\pm}(Q^2)=1-\frac{g_{KK\rho}}{g_{\gamma \rho}} \frac{Q^2}{Q^2+m_{\rho}^2}-
\frac{g_{KK\omega}}{g_{\gamma \omega}} \frac{Q^2}{Q^2+m_{\omega}^2}
-\frac{g_{KK\phi}}{g_{\gamma \phi}} \frac{Q^2}{Q^2+m_{\phi}^2}
\ee
with the coupling constants 
\be
   \frac{g_{KK \rho}}{g_{\gamma \rho}}=\frac{1}{2}\frac{g}{g_{\gamma}}, \quad 
\frac{g_{KK \omega}}{g_{\gamma \omega}}=\frac{1}{6}\frac{g}{g_{\gamma}}, \quad
\frac{g_{KK \phi}}{g_{\gamma \phi}}=\frac{\sqrt{2}}{6}\frac{g}{g_{\gamma}},
\ee
where $g/g_{\gamma}= 1.2$. 
Assuming equal masses for $\rho$ and $\omega$ ($m_V\equiv m_{\omega}=m_{\rho}$), one gets 
\be
F_{K^\pm}(Q^2)=1-\frac{g}{g_\gamma}\left(\frac{2}{3}\frac{Q^2}{Q^2+m_{V}^2}+
\frac{\sqrt{2}}{6} \frac{Q^2}{Q^2+m_{\phi}^2}\right) .
\ee
Hence, for low $Q^2$, the momentum dependence of the kaon form factor differs from the pion form factor by 
a factor $2/3+m_V^2/m_{\phi}^2\sqrt{2}/6\approx 0.8$. The mean square charge radius of the kaon is given as 
\be
\langle r_{K^\pm}^2 \rangle = \frac{g}{g_{\gamma
\rho}}\left(\frac{4}{m_{V}^2} 
+\frac{\sqrt{2}}{m_{\phi}^2}\right)=(0.32+0.06) \, \rm fm^{2}=0.38 \, \rm fm^2
\ee
as compared to $\langle r_{K^\pm}^2 \rangle =(0.34 \pm 0.05) \, \rm fm^2$, from
experiment \cite{amen86}. 
Hence, the contribution from the $\phi$ is small, but not negligible. 
If one takes into account only the $\rho$-contribution, then 
$\langle r_K^2\rangle=1/2 \langle r_{\pi}^2 \rangle =0.24 \, \rm fm^2$, 
which disagrees with the 
experimental value. 
For the form factor of the K$^0$, the contribution coming from the $\rho$-meson
changes its sign and one gets
\begin{equation}
\langle r_{K^0}^2 \rangle = \frac{g}{g_{\gamma
\rho}}\left(-\frac{2}{m_{V}^2} 
+\frac{\sqrt{2}}{m_{\phi}^2}\right)=-0.10 \mbox{ fm}^2
\end{equation}
which is again in the range of the experimental value of 
$\langle r_{K^0}^2\rangle= -0.054\pm 0.101$ fm$^2$ \cite{amen86}.  

The electromagnetic form factors of the baryons will be discussed in a
forthcoming publication \cite{becki}. 
\subsection{Meson-meson interaction}
\subsubsection{Vector meson masses}
Here we discuss the mass terms of the vector  mesons. 
The simplest scale-invariant 
form
\be
\label{vecfree}
{\cal L}_{vec}^{(1)}= \frac{1}{2} m_V^2 \frac{\chi^2}{\chi_0^2} \Tr V_{\mu} 
V^{\mu} +2g_4^4 \Tr (V_\mu V^\mu)^2
\ee
implies a mass degeneracy for the meson nonet. The first term of (\ref{vecfree}) 
is made scale invariant by multiplying it with 
an appropriate power of the glueball field $\chi$  (see Sec. \ref{scale} for 
details).
To split the masses, one can 
add the chiral invariant \cite{gasi69,mitt68}
\be
\label{lvecren}
{\cal L}_{vec}^{(2)} = \frac{1}{4} \mu \Tr\left[V_{\mu \nu} V^{\mu \nu} 
X^2 \right].
\ee  
Combining this with the kinetic energy term (Equation (\ref{kinetic})),
one obtains the following terms for the different vector mesons
\bea
\label{kinren}
 &-&\frac{1}{4} \left[1-\mu \frac{\sigma^2}{2}\right] (V_{\rho}^{\mu 
\nu})^2
 -\frac{1}{4} \left[1-\frac{1}{2} \mu (\frac{\sigma^2}{2}+\zeta^2)\right] 
   (V_{K^{\ast}}^{\mu \nu})^2 \\ \no
 &-&\frac{1}{4} \left[1-\mu \frac{\sigma^2}{2}\right](V_{\omega}^{\mu \nu})^2
 -\frac{1}{4} \left[1- \mu \zeta^2 \right] (V_{\phi}^{\mu \nu})^2 .
\eea 
The coefficients are no longer unity, therefore the vector meson fields have 
to be renormalized, i.e., the new $\omega$-field reads 
$\omega_r = Z_{\omega}^{-1/2} \omega$.
The renormalization constants are the coefficients  in the square 
brackets in front of the kinetic energy terms of Eq. (\ref{kinren}), 
i.e., 
$Z_{\omega}^{-1} = 1-\mu \sigma^2/2$. The mass terms of the vector mesons 
deviate from the mean mass $m_V $ by the renormalization 
factor\footnote{One could also split the $\rho-\omega$ mass degeneracy by adding 
a term of the form \cite{gasi69} 
$ (\Tr V_{\mu \nu})^2$ to Equation (\ref{kinren}). Or, alternatively, one could 
break the SU(2) symmetry of the vacuum allowing for a nonvanishing vacuum 
expectation value of the scalar isovector field. 
However,  the $\rho-\omega$ mass splitting is small 
($\sim$ 2 \%), 
and, therefore, we will not consider these complications.}, i.e., 
\be
m_{\omega}^2 = m_{\rho}^2=Z_{\omega} m_V^2 \quad ; \quad   
m_{K^{\ast}}^2 = Z_{K^{\ast}} m_V^2 \quad ; \quad 
m_{\phi}^2 = Z_{\phi} m_V^2 .
\ee
The constants $m_V$ and $\mu$ are fixed to give the correct $\omega$-and 
$\phi$-masses. The other 
vector meson masses are given in Tab. \ref{parameter}.

The axial vector meson masses can be described by adding 
terms analogous to (\ref{lvecren}). We refrain 
from discussing them further (see \cite{gasi69,ko94}).

\subsubsection{Scalar mesons}
The nonlinear realization of chiral symmetry offers many  more possibilities 
to form chiral invariants: the couplings of scalar 
mesons with each other are only governed by SU(3)$_V$-symmetry. 
However, only three kinds of independent invariants exist, namely
\be
\label{basis}
 I_1=\Tr X, \quad   I_2= \Tr X^2, \quad I_3=\det X .
\ee
All other invariants, $\Tr X^n$, with $n\ge3$, can be 
expressed as a function 
of the four invariants shown in (\ref{basis}). This can be shown from 
the characteristic equation of an arbitrary $3\times 3$ matrix $X$, 
\be
   (X-x_1)(X-x_2)(X-x_3)=0,
\ee  
where $x_i$ are the eigenvalues of $X$. By writing the coefficients of 
the powers of $X$ in terms of invariants one obtains
\be
\label{cole}
  X^3-I_1 X^2-\frac{1}{2}\left[I_2-(I_1)^2\right] X-I_3=0. 
\ee
Hence, 
one obtains the invariant $\Tr X^3$ as a function of the base  (\ref{basis}), 
\be
   I_{3m}\equiv\Tr X^3=I_1I_2+\frac{1}{2}\left[I_2-(I_1)^2\right]I_1+I_3.
\ee
By multiplying Equation (\ref{cole}) with, e.g., $X$ and taking the trace, 
the invariant for $n=4$ can be written in terms of Equation (\ref{basis}):
\be
 I_4\equiv\Tr X^4=I_1 I_{3m}+\frac{1}{2}\left[I_2 -(I_1)^2 \right] I_2+I_3 I_1 . 
\ee
A similar expression can be found for all other $n$.
Alternatively, instead of $I_3=\det X$ the invariant $I_{3m}=\Tr X^3$,  
can be chosen as an element of the basis. Then, $I_3$ can be rewritten in terms 
of the new basis $I_1$, $I_2$, and $I_{3m}$ as 
\be
   I_3 = \frac{1}{3} I_{3m}-\frac{1}{2} I_1 I_2+\frac{1}{6} (I_1)^3 .
\ee
For our calculations, the invariants of (\ref{basis}) are considered as 
building blocks, from which the different forms of the meson-meson interaction
are constructed. They
will be investigated including sets in which the models \cite{heid94} and 
\cite{tang95} are embedded in a chiral SU(3) framework 
(see Sec. \ref{work}).

\subsubsection{Broken scale invariance}
\label{scale}
The concept of broken scale invariance leading to the trace anomaly in
(massless) QCD, 
$\theta_{\mu}^{\mu}= \frac{ \beta_{QCD} }{2 g} {\cal G}_{\mu \nu}^a {\cal 
G}^{\mu \nu}_a$ (${\cal G}_{\mu \nu}$
is the gluon field strength tensor of QCD),
can be mimicked in an effective Lagrangian at tree level \cite{sche80}
through the introduction of the potential
\be
\label{lscale}
  {\cal L}_{\mathrm{scale}}=- k_4 \chi^4 - \frac{1}{4}\chi^4 \ln \frac{ \chi^4 }{ \chi_0^4}
 +\frac{\delta}{3}\chi^4 \ln \frac{I_3}{\det \langle X \rangle} .
\ee
The effect\footnote{According to \cite{sche80}, the 
argument 
of the logarithm has to be chirally and parity invariant. This is fulfilled by 
the dilaton, $\chi$, which is 
both a chiral singlet as well as a scalar.} of the logarithmic term $ \sim \chi^4 \ln 
\chi$ is to break the scale invariance. 
This leads to the proportionality $\theta_{\mu}^{\mu} \sim \chi^4$, 
as can be seen from  
\be
\theta_{\mu}^{\mu} = 4 {\cal{L}}  -\chi \frac{\partial {\cal L}}{\partial 
\chi}
- 2 \partial_{\mu} \chi \frac{\partial {\cal L}}
{\partial(\partial_{\mu} \chi)} = \chi^4  , 
\ee
which is a consequence of the definition of the scale transformations 
\cite{sche71}. 
This holds only, if the meson-meson potential is scale invariant. This can be achieved 
by multiplying the invariants of scale dimension less then four 
with an appropriate power of the dilaton 
field $\chi$.

The comparison of the trace anomaly of 
QCD with that of the effective theory allows 
for the identification of the $\chi$-field with the gluon condensate: 
\be
\theta_{\mu}^{\mu} =  \left\langle \frac{ \beta_{QCD} }{2 g} {\cal G}_{\mu 
\nu}^a 
{\cal G}^{\mu \nu}_a \right\rangle
 \equiv  (1-\delta)\chi^4 .
\ee
The parameter $\delta$ originates from the second logarithmic term with the 
chiral invariant $I_3$ (see also \cite{heid94} for the 
chiral SU(2) linear $\sigma$-model). 
An orientation 
for the value of $\delta$ may be taken from $\beta_{QCD}$ at the one loop level, 
with $N_c$ colors and $N_f$ flavors, 
\be
\label{qcdbeta}
   \beta_{QCD}=-\frac{11 N_c g^3}{48 \pi^2} \left(1-\frac{2N_f}{11 N_c}\right)
   +{\cal O}(g^5) . 
\ee                               
Here the first number in parentheses arises from the (antiscreening) 
self-interaction of the gluons and the second term, proportional to $N_f$, 
is the (screening) contribution of quark pairs. Equation (\ref{qcdbeta}) suggests 
the value $\delta=6/33$ for three flavors and three colors. This value 
gives the order of magnitude about which the parameter $\delta$ will be 
varied.

For simplicity, we will also consider the case in which $\chi=\chi_0$, 
where the gluon condensate does not vary with density. We will refer to this 
case as the {\it frozen glueball limit}.

\subsection{Explicitly broken chiral symmetry}
\label{esb}
In order to eliminate the Goldstone modes from a chiral effective theory, 
explicit symmetry breaking terms have to be introduced. 
Here, we use 
\be
\label{esb-gl}
 {\cal L}_{SB}  =  -\frac{1}{2} m_{\eta_0}^2\Tr Y^2 
     -\frac{1}{2} \Tr A_p \left(uXu+u^{\dagger}Xu^{\dagger}\right)
     -\Tr \left(A_s-A_p\right) X . 
\ee
The first term, which breaks the $U(1)_A$ symmetry, gives a mass to the pseudoscalar singlet. 
The second term is motivated by the explicit symmetry breaking  
term of the linear $\sigma$-model, 
\be
\label{lesb}
   \frac{1}{2}\Tr A_p(M+M^\dagger)=\Tr A_p\left(u(X+iY)u+u^\dagger(X-iY)u^\dagger\right) , 
\ee
with $A_p=1/\sqrt{2}{\mathrm{diag}}(m_{\pi}^2 f_{\pi},m_\pi^2 f_\pi, 2 m_K^2 f_K
-m_{\pi}^2 f_\pi)$ and $m_{\pi}=139$ MeV, $m_K=498$ MeV. 
For simplicity, $\eta_0/\eta_8$ mixing is neglected 
 by  omitting $Y$ from the second term of equation
(\ref{esb-gl}). If this term is included, we get a mixing angle 
of $\theta=16^{o}$ for parameter set $C_1$ (see section \ref{szint}),
which agrees well with experiment, $\theta^{\rm{exp}} \approx
20^{o}$ from $\eta,\eta' \to \gamma \gamma$. 
 
In the 
case of $SU(3)_V$-symmetry, the quadratic Gell-Mann Okubo mass formula, 
$3 m_{\eta_8}^2+m_{\pi}^2-4m_K^2=0$, is satisfied.

The third term breaks  SU(3)$_V$-symmetry. $A_s={\mathrm{diag}}(x,x,y)$ 
can be used to remove the vacuum constraints on the parameters of the meson-meson potential 
by adjusting $x$ and $y$ in such a way that the terms linear in $\sigma$ and $\zeta$ vanish in the vacuum.\\

\section{Mean-field approximation}
\label{mfa}
The terms discussed so far involve the full quantum operator fields which cannot 
be treated exactly. To apply the model to the description of finite nuclei, 
we perform the mean-field approximation. This is a nonperturbative relativistic 
method to solve approximately the nuclear many body problem by replacing the 
quantum field operators by its classical expectation values (for a recent review 
see \cite{serot97}). 

In the following, we will consider the time-independent, spherically symmetric case
of finite nuclei with vanishing net strangeness, i.e. only nucleons, and zero temperature. 
As usual, only the time-like component of the vector mesons
$\omega \equiv \langle \omega_0 \rangle$ and 
$\rho \equiv \langle \rho_0 \rangle$ 
survive in the mean-field approximation.
Additionally, due to parity conservation we have $\langle \pi_i \rangle=0$.
The strange vector field $\phi$ does not couple to the nucleon. Therefore, for simplicity 
it is omitted in the mean-field version of the Lagrangian (\ref{lagrange}), which reads: 
\begin{eqnarray}
\label{mfala}
{\cal L}_{kin} &=& -i\overline{N} \gamma_i \nabla^i N 
- \frac{1}{2} \sum_{\varphi=\sigma, \zeta,\chi,\omega,\rho, A}
\nabla_i \varphi \nabla^i \varphi\\
{\cal L}_{BM}+{\cal L}_{BV} &=& -\overline{N}\gamma_0\left[ g_{N \omega} \omega_0 
+g_{N \rho}\tau_3 \rho_0+\frac{1}{2}e(1+\tau_3)A_0 +m_N^{\ast} \gamma_0\right]N \no \\ 
{\cal L}_{vec} &=& \frac{ 1 }{ 2 } \frac{\chi^2}{\chi_0^2}
\left(m_{\omega}^{2}\omega^2  +m_{\rho}^{2}\rho^2\right)
+g_4^4 \left({\omega}^{4}+6 \omega^2 \rho^2 +\rho^4\right)  \no \\
{\cal L}_0 &=& -\frac{ 1 }{ 2 } k_0 \chi^2 (\sigma^2+\zeta^2) 
+ k_1 (\sigma^2+\zeta^2)^2 
     + k_2 \left( \frac{ \sigma^4}{ 2 } + \zeta^4\right) 
     + k_3 \chi \sigma^2 \zeta \no \\
&+&k_{3m} \chi\left(\frac{\sigma^3}{\sqrt{2}}+\zeta^3\right)
- k_4 \chi^4 - \frac{1}{4}\chi^4 \ln \frac{ \chi^4 }{ \chi_0^4}
 +\frac{\delta}{3}\chi^4\ln \frac{\sigma^2\zeta}{\sigma_0^2 \zeta_0} \no  \\
{\cal L}_{SB} &=& -\left(\frac{\chi}{\chi_0}\right)^{2}\left[x 
\sigma + y\zeta \right] . \no 
\end{eqnarray}
Equation (\ref{mfala}) is the most general mean-field Lagrangian within our discussion 
of which different subsets 
of parameters and terms are discussed in Sec. \ref{work}.

From the Lagrangian (\ref{lagrange}), the following equations of motion for the 
various fields are derived:
\begin{eqnarray}
\label{eom}
D \omega &=& -\omega  \left(\frac{\chi}{\chi_0}\right)^2 m_{\omega}^2 \omega
          -4\,g_4^4 (\omega^3+3 \rho^2 \omega)+g_{\omega N} \rho_B \\
D \rho &=&   -\rho \left(\frac{\chi}{\chi_0}\right)^2 m_{\rho}^2 \rho
           -4\,g_4^4(\rho^3+3\rho \omega^2) +g_{\rho N} \rho_B \no \\
D \chi &=& -\frac{\chi}{\chi_0^2}\left( m_{\omega}^2\omega^2  +m_{\rho}^2\rho^2\right)
        + k_0 \chi (\sigma^2+\zeta^2) 
        - k_3 \sigma^2 \zeta -k_{3m}\left(\frac{\sigma^3}{\sqrt{2}}+\zeta^3\right) \no \\
          &+& \left( 4 k_4 + 1 + 4 \ln \frac{ \chi }{\chi_0}
        - 4 \frac{\delta}{3} \ln \frac{\sigma^2 \zeta}{\sigma_0^2\zeta_0}
 \right) \chi^3 \nonumber+ 2\frac{\chi}{\chi_0^2} \left[x\sigma+y\zeta\right] \no \\  
D \sigma &=& k_0 \chi^2 \sigma - 4 k_1 (\sigma^2+\zeta^2)\sigma 
 - 2k_2 \sigma^3 
          - 2 k_3 \chi \sigma\zeta -3k_{3m}\chi \frac{\sigma^2}{\sqrt{2}}
-\frac{2\delta \chi^4}{3 \sigma}  + \left(\frac{\chi}{\chi_0}\right)^{2} x
+ \frac{\partial m_N^{\ast}}{\partial \sigma} \rho_s \no \\                 
D \zeta &=& k_0 \chi^2 \zeta - 4 k_1 (\sigma^2+\zeta^2) \zeta 
- 4 k_2 \zeta^3 - k_3 \chi \sigma^2 -3 k_{3m}\chi\zeta^2  
-\frac{\delta \chi^4}{3\zeta}
 + \left(\frac{\chi}{\chi_0}\right)^{2} y+\frac{\partial m_N^{\ast}}{\partial \zeta}  \rho_s \no
\end{eqnarray}
The Dirac equation for the nucleon and the equation for the photon field 
are of the form given, e.g. by Reinhard \cite{Rei89a} and need not to be repeated here .
The densities $\rho_s=\langle \ovl{N}N \rangle$, $\rho_B=\langle \ovl{N}\gamma_0N \rangle$, 
$\rho_3=\langle \ovl{N}\gamma_0 \tau_3 N \rangle$ can be expressed in terms of 
the components of the nucleon Dirac spinors in the usual way \cite{serot97}. 
In equations (\ref{eom}),
the spatial derivatives are abbreviated by $D\equiv
-\nabla^2-\frac{2}{r}\nabla$.

The set of coupled equations are solved using an accelerated gradient
iteration method
following \cite{rufa88}. The Dirac equation for the nucleons can be cast in a
modified Schr\"odinger equation with an effective mass. The meson field
equations reduce to radial Laplace equations. 
In each iteration step, 
the coupled equations for the nuclear radial wave functions are solved for the
given potentials, the corresponding densities are calculated, 
then the meson field equations are solved for the given densities, so that the new
potentials are derived and the next iteration step can begin until convergence is
achieved. The meson field equations are solved in the form
\begin{equation}
\left[ -\frac{d^2}{dr^2} + m^2_{\varphi,0} \right] (r\varphi^{(N+1)}) = 
-r f(\rho,\varphi^{(N)})
\end{equation}
where $m^2_{\varphi,0}$ is the vacuum mass of the respective meson (or an
arbitrary mass) which is
subtracted on the right hand side of the equation. The
function $f(\rho,\varphi^{N})$ stands for the interaction terms with other
meson fields, the source
terms coming from the nucleon density and the selfinteraction terms as given above.
This form achieves a five-point precision for the Laplacian by using only a
three-point formula by solving for $(r\varphi)$. The scalar fields have to be
solved by replacing e.g.\ $\sigma \to (1-\sigma/\sigma_0)$ to ensure the boundary
condition that the field has to vanish for $r\to\infty$. The iteration is
damped by taking into account only a fraction of the newly calculated density
for the next iteration step. 

The energy-momentum tensor can be used to obtain the total energy of the 
system in the standard way \cite{serot97}. After eliminating the gradient terms on 
the fields by using the field equations, one obtains
\be
   E=\sum_{\alpha}^{\rm{occ}} \epsilon_{\alpha} (2 j_{\alpha}+1)-\frac{1}{2} \int
dr r^2 (m_N^{\ast} \rho_s+g_{N \omega} \omega\rho_B+g_{N \rho} \rho \, \rho_3)+E_{\rm{rearr}}
\ee
In the first term $\epsilon_{\alpha}$ are the Dirac single particle energies 
and $j_{\alpha}$ is the total angular momentum of the single particle state. 
In nuclear matter this term becomes $4 \sum_k \left( g_\omega \omega_0+\sqrt{k^2+m^{\ast 2}}\right)$.
The rearrangement energy $E_{\rm{rearr}}$ is 
\bea
\label{rearr}
   E_{\rm{rearr}} &=&\int dr r^2 \left(  2 g_4^4\left(\omega^4+\rho^4+6\omega^2 \rho^2+2\phi^4\right)
    - 2 k_1\left(\sigma^2+\zeta^2 \right)^2
    - 2 k_2 \left(\frac{\sigma^4}{2}+\zeta^4\right) \right.\\ \no
    &-& \left. k_3 \chi_0\sigma^2 \zeta-k_{3m}\chi_0 \left(\frac{\sigma^3}{\sqrt{2}}+\zeta^3\right)
           +2\frac{\delta}{3}\chi_0^4 \ln\left(\frac{\sigma^2 \zeta}{\sigma_0^2 \zeta_0}\right)
            -x\sigma-y\zeta   \right)-V_{\rm{vac}}
\eea
The constant $V_{\rm{vac}}$ is the vacuum energy which is subtracted to yield zero energy in 
the vacuum. Equation (\ref{rearr}) is the rearrangement energy for the frozen glueball 
model which is used for most of the fits discussed in the following. 
Let us now proceed to study the application to physical hadrons and hadronic 
matter fits.
\section{Chiral models that work}
\label{work}
As was pointed out in \cite{furn93}, reproducing the nuclear matter 
equilibrium point is not sufficient to ensure a quantitative 
description of nuclear phenomenology. For this, one has to study the 
systematics of finite nuclei. 
This is done in the following for various potentials in a chiral SU(3) 
framework. Those include the potential of 
the SU(3) linear $\sigma$ model, the potential of the Minnesota-group \cite{heid94}
and the Walecka model including nonlinear cubic and quartic self-interactions
of the scalar field \cite{boguta77,fpw}. 
\subsection{Potential of the linear $\sigma$-model}
\label{szint}
The potential of the linear $\sigma$-model is particularly interesting because 
the strange condensate couples to the nonstrange 
condensate,  $\sigma$,  in such a way that it deviates from its VEV even in the 
case of a system containing only nucleons. 
With the scale breaking logarithm included 
(${\cal L}_{\mathrm{scale}}$, see (\ref{lscale})), it reads
\bea 
\label{cpot}
{\cal L}_0^C&= &  -\frac{ 1 }{ 2 } k_0 \chi^2 I_2
     + k_1 (I_2)^2 + k_2 I_4 +2 k_3 \chi I_3+{\cal L}_{\mathrm{scale}}.
\eea
Here, the explicit symmetry breaking term of the linear $\sigma$-model 
is used, i.e., $A_s=A_p$, which implies the same term to break the chiral 
symmetry 
in the scalar and pseudoscalar sector, respectively. 
In addition,  the mass term of the pseudoscalar 
singlet is set to 
\be
\label{etamas}
    m_{\eta_0}^2=k_0 \chi_0^2-4\left(\frac{k_2}{3}+k_1\right)\left(\sigma_0^2+\zeta_0^2\right)
+\frac{4}{3}k_3\chi_0\left(\zeta_0+\sqrt{2}\sigma_0\right)-\frac{4}{9}\delta \chi_0^4\left(\frac{1}{\sigma_0^2}+\frac{\sqrt{2}}{\sigma_0 \zeta_0}\right), 
\ee
This is equal to the pseudoscalar singlet mass which is obtained 
if $M$ and $M^\dagger$ 
of the linear $\sigma$-model potential \cite{paper2} are replaced by 
Eqs. (\ref{mmdagger}).

The elements of the matrix $A_p$ are fixed to fulfil 
the PCAC-relations of the pion and the kaon, respectively. 
Therefore, the parameters of the chiral invariant potential, 
$k_0$ and $k_2$, are used to ensure an extremum in the vacuum. 
As for the remaining constants, 
$k_3$ is constrained by the $\eta'$-mass, and $k_1$ is varied 
to give a $\sigma$-mass of the order of $m_{\sigma}=500$ MeV. The VEV of the 
gluon condensate, $\chi_0$, is fixed to fit
the binding energy of nuclear matter $\epsilon_0/\rho-m_N=-16$ MeV at the 
saturation density $\rho_0 =0.15 \, \rm fm^{-3}$. The VEV of the fields 
$\sigma_0$ 
and $\zeta_0$ are constrained by the decay constants of the 
pion and the kaon, respectively (see Eq. (\ref{zeta0})).

With the same potential, equation (\ref{cpot}), fits with ($C_1$) and without ($C_2$)
 a dependence 
of the nucleon mass on the strange condensate $\zeta$ can be done. 
To see, whether there is 
a significant effect from the gluon condensate $\chi$ at moderate densities, 
a nonfrozen 
fit is also studied ($C_3$) where we allow the condensate of the dilaton
field to deviate from its vacuum value.

As can be seen from Tab. \ref{massen}, 
the hadronic masses in the vacuum  have reasonable values. 
If the potential of Eq. (\ref{cpot}) in combination with (\ref{etamas}) 
is used, the mass
of the $\eta'$ meson depends on all constants $k_i$ and on $\chi_0$, which 
are also used to fit nuclear matter properties. In our fits, the pseudoscalar 
meson masses have the values $m_{\eta}=574$ MeV and $m_{\eta'}=969$ MeV.

According to Tab. \ref{nucmat}, the values of the effective nucleon mass 
and the compressibility in the medium (at $\rho_0$) are reasonable. 
For a fine-tuning of the single particle energy levels and a lowering of the 
effective nucleon mass, a quartic term for vector mesons (see Equation (\ref{vecfree})) has 
to be taken into account. 

Once the parameters have been fixed to nuclear matter at $\rho_0$ the
condensates and hadron masses at high baryon densities can be
investigated. 

\begin{figure}
\epsfbox{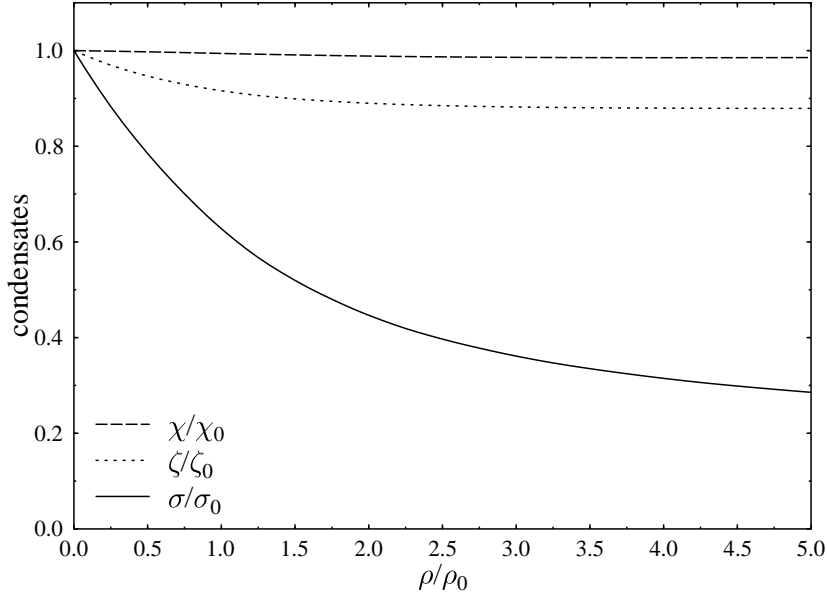}
\caption{\label{felder} Scalar condensates $\sigma$, $\zeta$ 
and $\chi$ as a function of the baryon density for zero 
net strangeness.}
\end{figure}

In Fig. \ref{felder} we display the scalar mean fields
$\sigma$, $\zeta$ and $\chi$ as a function of the baryon density for
vanishing strangeness. One sees that the gluon condensate 
$\chi$ stays nearly constant when the density is raised, so that
the approximation of a frozen glueball is reasonable.
The strange condensate $\zeta$ is only reduced by about 10 percent from
its vacuum expectation value. This is not surprising since there are only 
nucleons in the system and the nucleon--$\zeta$ coupling is fairly weak.
The main effect occurs for the non--strange condensate $\sigma$. The 
field has dropped to 30 percent of its vacuum expectation value at 
4 times normal nuclear density. If we extrapolate to even higher densities
one observes that the $\sigma$ field does not change significantly, so for 
all fields a kind of saturation takes place at higher densities.

From equation (\ref{bmver1}) one sees that the baryon masses are generated
by the non--strange condensate $\sigma$ and the strange condensate $\zeta$.
So the change of these scalar fields causes the change of the
baryon masses in medium.

\begin{figure}
\epsfbox{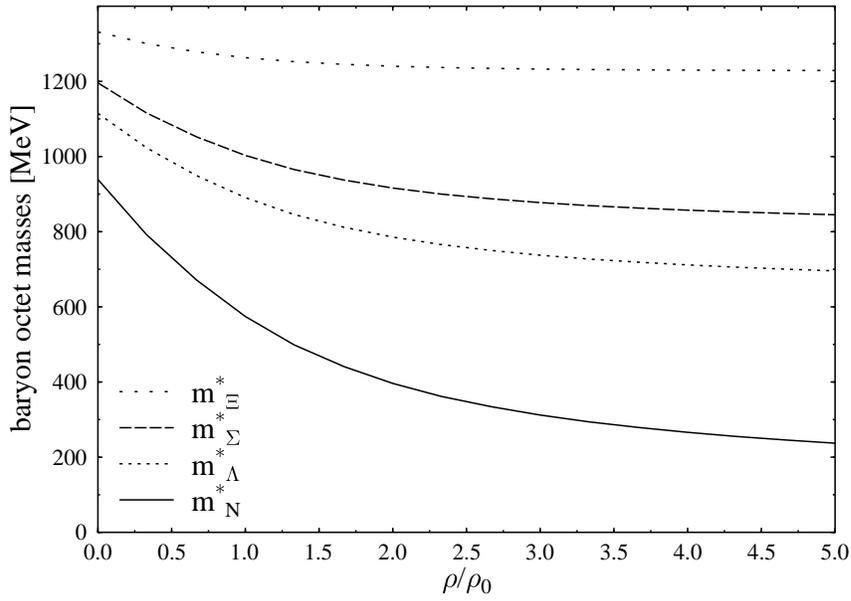}
\caption{\label{bmassen} Effective baryon
masses as a function of the baryon density for zero 
net strangeness.}
\end{figure}

The density dependence of the effective baryon masses $m_i^{\ast}$ is shown in
Fig. \ref{bmassen}. When the density in the system is raised, the masses 
drop significantly up to $4$ times normal nuclear density.
This corresponds to the above mentioned behavior of the condensates. 
Furthermore, one  observes that the change of the baryon mass differs with 
the strange quark content of the baryon. This is caused by
the different behavior of the non--strange condensate $\sigma$ which 
mainly couples to the nonstrange part of the baryons, and 
the strange condensate $\zeta$ which couples mainly to the strange part of the
baryons.


Without changing the parameters of the model, 
the properties of nuclei can be predicted readily.

\begin{figure}
\epsfbox{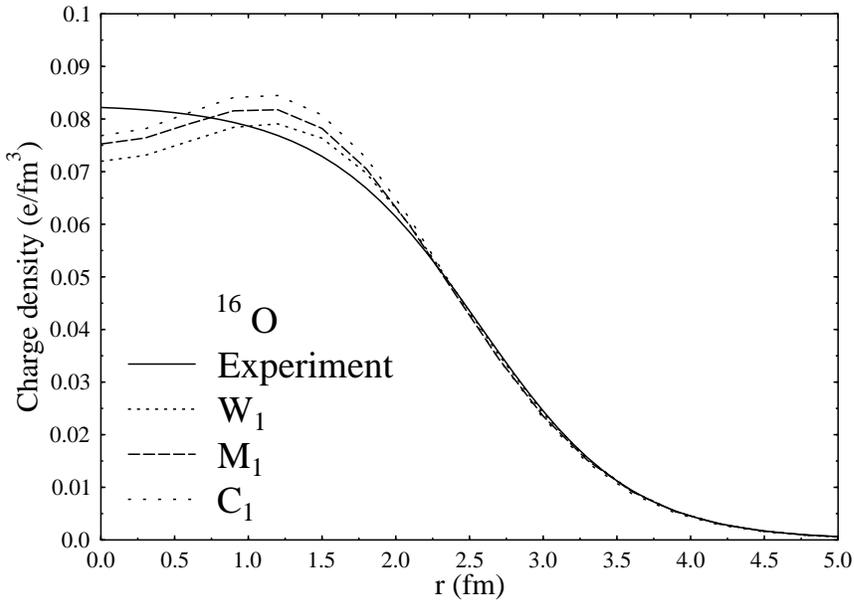}
\caption{\label{rcho} Charge density for $^{16}$O for the parameter
sets indicated. The experimental charge density is fitted 
with a three-parameter Fermi model \protect\cite{vries87}.}
\end{figure}

\begin{figure}
\epsfbox{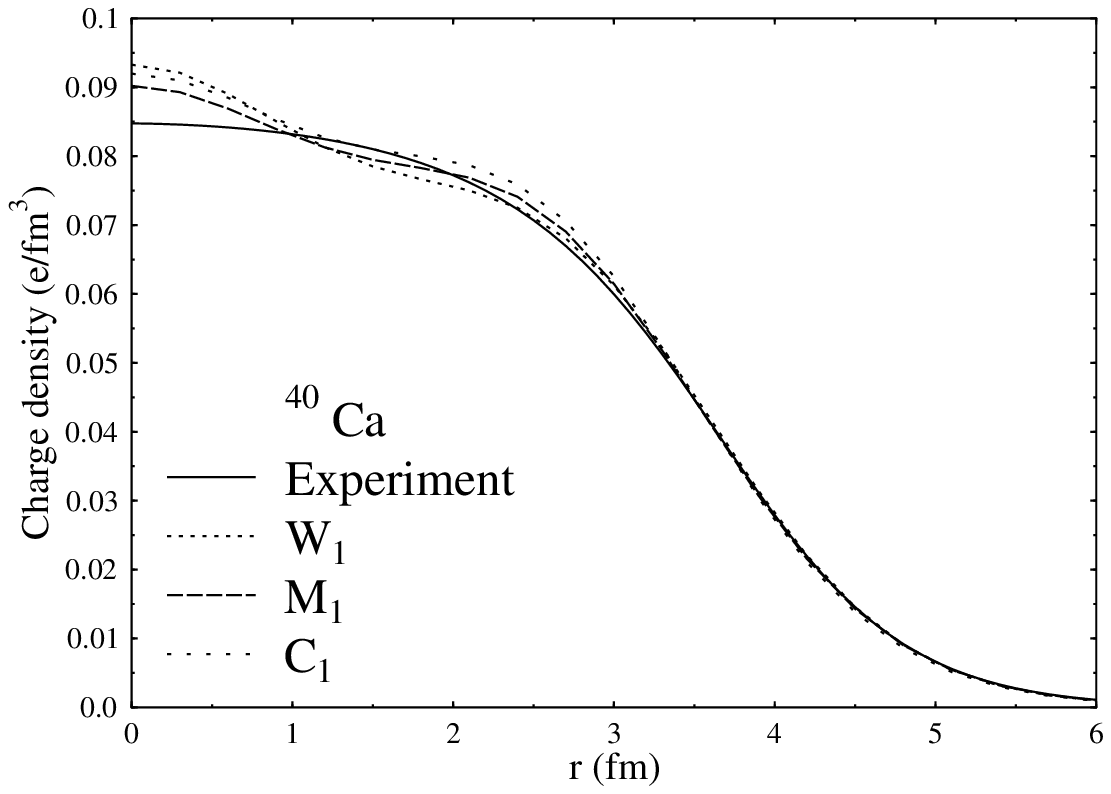}
\caption{\label{rchca}As for Fig. \ref{rcho}, but for $^{40}$C.}
\end{figure}

\begin{figure}
\epsfbox{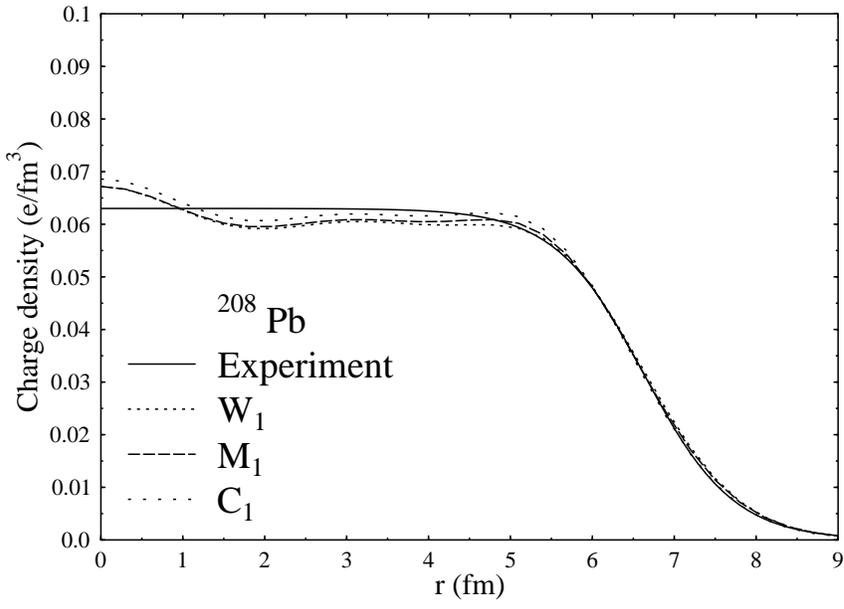}
\caption{\label{dichten}As for Fig. \ref{rcho}, but for $^{208}$Pb.}
\end{figure}

\begin{figure}
\epsfbox{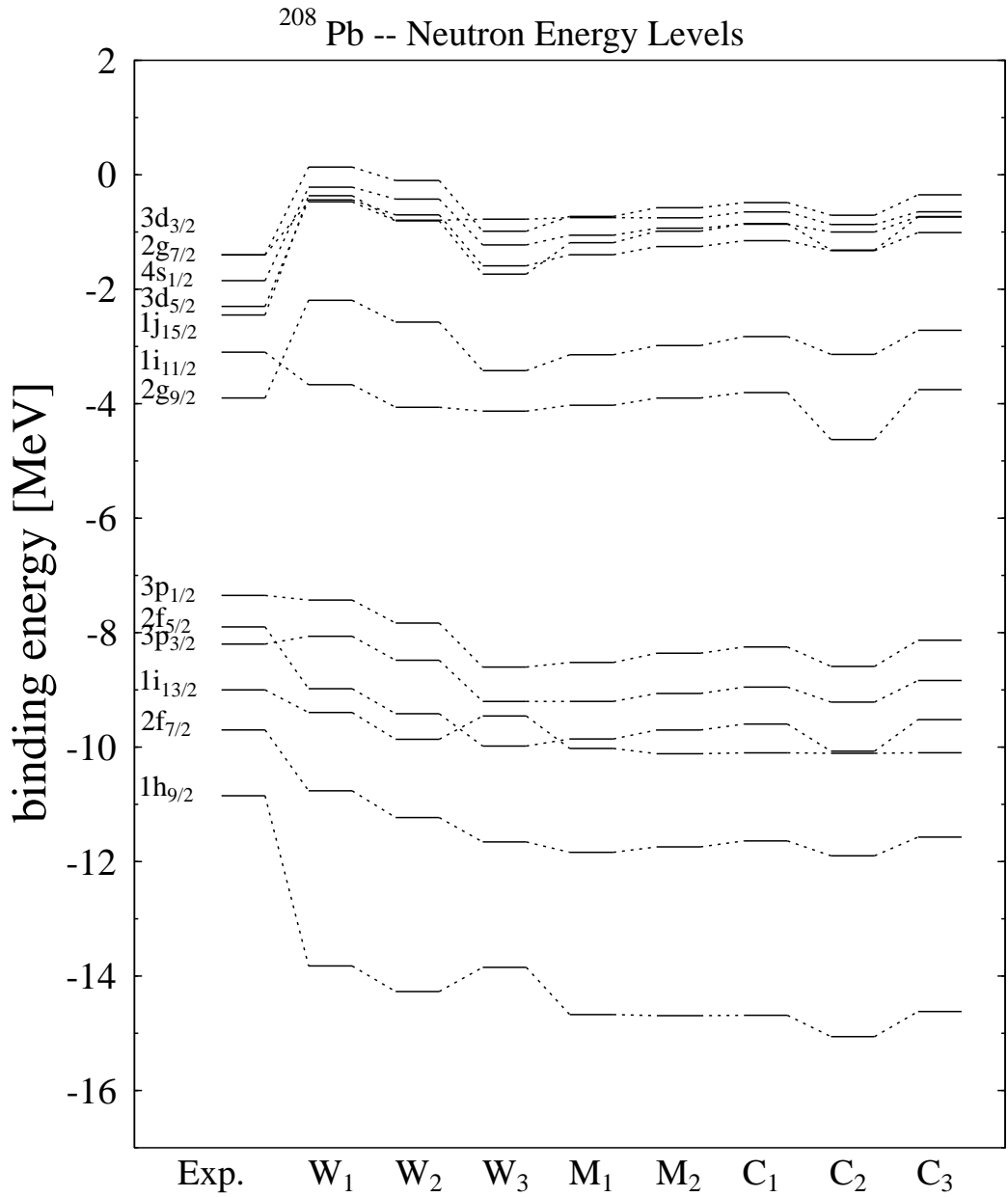}
\caption
{\label{spe-n}Single particle energies of neutrons near the Fermi
energy in $^{208}$ Pb. Experimentally measured levels are compared with
predictions from various potentials used (see text).}
\end{figure}

\begin{figure}
\epsfbox{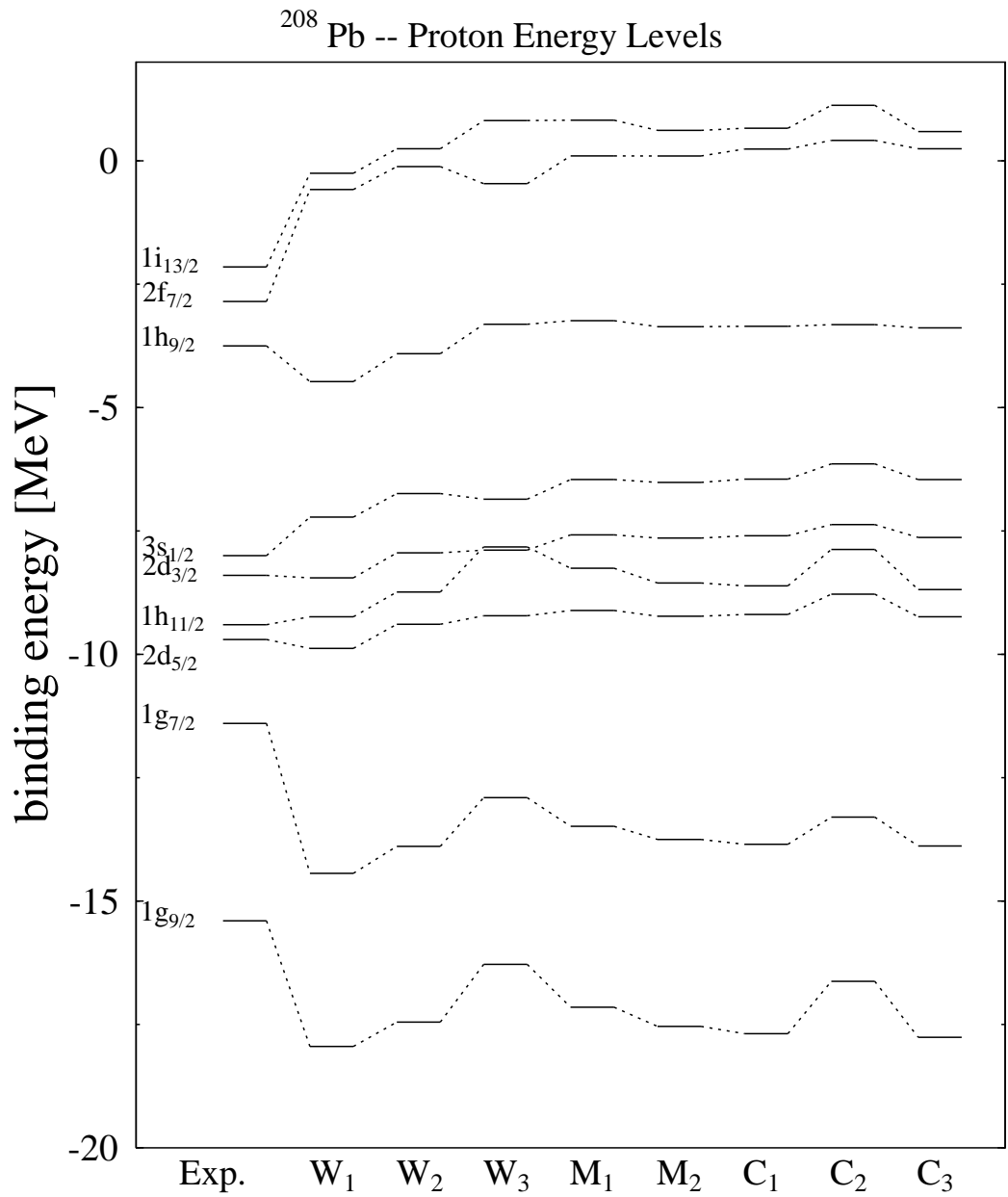}
\caption{\label{spe-p}As for Fig. \ref{spe-n}, but for protons.}
\end{figure}

The charge densities of $^{16}$O, $^{40}$Ca and $^{208}$Pb are found to have 
relatively small radial oscillations  (Figs. \ref{rcho}, \ref{rchca}, and \ref{dichten}), though 
such oscillations cannot be found in the data\footnote{Similar problems exist also for 
nonchiral models, for a discussion see \cite{furn93,price90}} The
experimental charge 
densities are taken from \cite{vries87}, 
where a three-parameter Fermi model was used to fit the data\footnote{A more 
sophisticated model-independent analysis by means of an expansion for the charge 
distribution as a sum of Gaussians would lead to an even closer correspondence between our 
results and the experimental data.}.
The charge radii are close to 
the experimental observation (Tab. \ref{nuclei}).
The binding energies of $^{16}$O, $^{40}$Ca and $^{208}$Pb are in reasonable 
agreement with the experimental data. Nevertheless they are low by
approximately 0.5 MeV. To correct 
this, a direct fit to nuclear properties has to be done \cite{becki}. 
As can be seen from Tab. \ref{nuclei}, models $C_1$ and $C_2$ exhibit a 
spin-orbit splitting that lies
 within the band of the experimental uncertainty given in \cite{rusnak}. 
The single-particle energies of $^{208}$Pb are close to those of the 
Walecka model extended to include nonlinear $\sigma^3$ and $\sigma^4$
terms \cite{fpw} or the model
\cite{heid94}, both for neutrons (Fig. \ref{spe-n})  and for protons 
(Fig. \ref{spe-p}).  This is encouraging since neither the
nucleon/scalar meson nor the nucleon/$\rho$ meson coupling constants can 
be adjusted to nuclear matter or nuclei properties, in contrast to the
Walecka model \cite{fpw}. 

\subsection{Minnesota model}
\label{minne}
By incorporating the physics of broken scale invariance in the form of 
a dilaton field and a logarithmic potential, the Minnesota group succeeded 
in formulating a model 
with equally good results as those of \cite{fpw}
in the context of a linearly realized symmetry \cite{heid94}.
When switching to SU(3), it is necessary to use a nonlinear 
realization, because there is no freedom in the linear representation to 
correct for the unrealistic hyperon potentials \cite{paper2} 
if one adopts a Yukawa-type baryon-meson interaction. 

With a potential of the form
\be
\label{minpot}
{\cal L}_0^{M} =  -\frac{ 1 }{ 2 } k_0 \chi^2 I_2+{\cal L}_{\mathrm{scale}}
\ee
the model \cite{heid94} is embedded in $SU(3)$. Those results can be reproduced exactly ($M_1$-fit). 
Here, the parameter for explicit symmetry breaking (cf Eq.(\ref{esb-gl})) 
$A_s={\rm{diag}}(0,0,y)$ is used, 
where $y$ is adjusted as to eliminate the terms linear in $\zeta$. 
For $y=0$, (or, generally, a matrix $A_s$ proportional to the unit matrix) 
the vacuum is SU(3)$_V$-invariant. 

Even with the SU(3)-constraint on the nucleon-$\rho$ coupling, $g_{N \rho}=g_{N \omega}/3$
and with a coupling of the strange condensate to the nucleon according to Eq. \ref{bmver1} (fit $M_2$), 
the results are of the same quality as those obtained in \cite{heid94}. 

Generally, the potential (\ref{minpot}), in which the two condensates 
$\sigma$ and $\zeta$ are decoupled from each other, leads to 
scalar masses which are all of the order of 500 MeV. To correct this 
failure, additional terms have to be included which lead to a 
$\sigma/\zeta$-mixing, as e.g., the linear $\sigma$ model potential
(see Sec. \ref{szint}).
%

\subsection{Chiral Walecka model} 
\label{wale}
As in the linear $\sigma$-model, the coupling 
constant of the nucleon to the $\sigma$ meson, $g_{N \sigma}$,  
is constrained to yield the correct nucleon mass, 
\be
 g_{N \sigma}=\frac{m_N}{f_{\pi}} .
\ee
To reproduce exactly the results obtained in the nonlinear
$\sigma-\omega$ model \cite{fpw}, it is necessary to keep this coupling 
as a free parameter. For that purpose, we introduce the additional 
term 
\be
   -m_{\mathrm{av}} \Tr \ovl{B} B  , 
\ee
which should be a small correction to the dynamically generated 
nucleon mass. 
In the nonlinear realization of chiral symmetry, this term is 
chirally invariant.

In order to obtain a chiral model which is capable of exactly 
reproducing the results of the nonlinear Walecka  model \cite{fpw}, 
it is necessary to include only terms in the meson-meson potential, 
in which both condensates, $\sigma$ and $\zeta$,  
are decoupled from each other:
\be
\label{walepot}
{\cal L}_0^{W} =  -\frac{ 1 }{ 2 } k_0 \chi^2 I_2
   +k_{3m} \chi I_{3m}  +k_2 I_4. 
\ee
Here, the scale breaking potential is neglected by taking the frozen 
glueball limit and setting $\delta=0$. 
To allow for a free adjustment of the parameters $k_0$, $k_{3m}$, and $k_4$ 
to nuclear matter properties, $A_s$ is set to 
\be
A_s=\mbox{diag}(x,x,y).
\ee 
With $x$ and $y$ 
one then has two additional parameters to eliminate linear fluctuations in $\sigma$
and $\zeta$. 
The symmetry in the scalar sector is only broken explicitly if 
$y\ne x$. 

The $\sigma$ field used here has a nonvanishing vacuum expectation value 
as a result of the spontaneous 
symmetry breaking\footnote{In \cite{tang97}, the results of the
Walecka model could also be reproduced in a nonlinear SU(2) chiral
approach. There, however, the limit $m_{\sigma} \to \infty$ has been
performed introducing in a second step a light scalar $\sigma$ field mimicking
correlated $2\pi$-exchange. 
In addition, the hadron masses were not generated dynamically.}. To 
compare this field $\sigma$ with the field $s$   used in 
the nonlinear Walecka model \cite{fpw}, one has to perform the transformation
\be
       \sigma=\sigma_0 +s . 
\ee
After inserting this transformation into the potential (\ref{mfala}), 
one can identify the parameters used here with those of \cite{fpw}, 
\bea
  m_{\sigma}^2&=&k_0 \chi_0^2-3 k_{3m} \sigma_0 \sqrt{2}-6 k_2 \sigma_0^2 \\
  \kappa  &=& -3 k_{3m} \sqrt{2}-12 k_2 \sigma_0 \\
  \lambda &=& -12 k_2
\eea
Therefore, the results obtained in the framework of the 
Walecka model \cite{fpw} can be reproduced 
exactly\footnote{For this, a small $d$-type admixture of the
baryon/vector-meson 
coupling is necessary, since the relation $g_{N\omega}=3 g_{N \rho}$
is not fulfiled exactly in the Walecka model. (For the set $W_1$,
$\alpha_v=0.95$ is used).} within this ansatz (from now 
one denoted $W_1$) given a special choice of explicit symmetry
breaking. However, in contrast to the Walecka model the hadron 
masses are generated spontaneously. 

The masses of the scalar multiplet as resulting from the parameterization of \cite{fpw}
are of the order of 500 MeV, as can be read off Tab. \ref{massen}. To correct for
 this, terms which mix the $\sigma$ with the $\zeta$ have 
to be added (see below).

A problem, which is well known in the context of the 
Boguta-Bodmer model,  
exists here, too: For certain combinations of parameters 
the potential is not bound from below. To cure 
this problem, one can introduce additional terms, as was done 
in \cite{pg}. Another, more physical, way to circumvent 
this problem is to use the physics of broken scale invariance, 
as in \cite{heid94} or the models used in Secs. \ref{szint} and \ref{minne}. 

Beyond exactly reproducing an existing, successful model,  
it is interesting to ask, whether improvements in the phenomenology
can be made as compared to the 
Walecka model. This could mean either reducing 
the amount of parameters needed, or a significantly improved 
description of 
existing data, or the description of a broader range of 
physical phenomena.
 
Let us first consider the limit $m_{\mathrm{av}}$=0. Then, the 
relation 
\be
\label{gns}
  g_{N \sigma}=\frac{m_N}{f_{\pi}} 
\ee
known from the linear $\sigma$-model is valid. To reproduce exactly  
the results of \cite{fpw} (Fit W1), $m_{av}=32$ MeV, which is about $3\%$
as compared to $g_{N \sigma}\sigma_0$ and which is roughly 
of the same order as the sum of the current quark masses in the baryon. 
Indeed, the model (fit $W_1$) 
does not give worse results than 
the model $W_2$ where the relation (\ref{gns}) 
and the SU(3)-symmetry constraint
$g_{N \omega}=3 g_{N \rho}$, corresponding to a value 
$a_v=1$, (Eq. (\ref{quarkcoupling})), is used. 

Next, it is desirable to have masses for the scalar nonet which are 
(except for 
$m_\sigma$) on the order of 1 GeV. This can be achieved by admitting 
 mixing between the 
$\sigma$ and the $\zeta$ by including the term 
$k_1 (I_2)^2$ to the scalar potential (\ref{walepot})
(fit $W_3$). Therefore, in the SU(3)-framework, 
even for a pure system of only nucleons it is {\it necessary} to take the strange 
condensate $\zeta$ into account! 

%

\subsection{Hyperon central potentials}
As discussed in the introduction, the reasons for unrealistic hyperon potentials in 
the linear $\sigma$-model are the different types of coupling 
of the spin-0 and spin-1 mesons to baryons and a direct coupling 
of the $\sigma$ with the strange condensate. The second reason produces too deep hyperon 
central potentials since the additional attraction 
stemming from the $\zeta$ cannot be compensated with an additional 
repulsion from the $\phi$. This has a vanishing expectation value in nuclear matter
 at zero net strangeness 
since it does neither couple to the nucleon nor to the $\omega$.

Both effects can be switched off in the nonlinear realization
(Fits $W_1$, $W_2$, $M_1$, and $M_2$). However, even in those fits, the experimentally 
extracted value for the $\Lambda$ central potential of $U_{\Lambda}=28 \pm 1$ MeV \cite{bodmer}
cannot be reproduced. The nucleon central potential of $U_N \approx -70$ MeV is too deep:
$\frac{2}{3}U_N \ne -28$ MeV. A shallower potential for the 
nucleon leads to too small a  spin-orbit splitting of the energy levels of nucleons. 
Therefore, both the central potentials 
of the nucleon and of the $\Lambda$ cannot be reproduced if the f-type, quark-model motivated,  
coupling constant is used for both baryon vector-meson  and baryon scalar-meson interaction.
The sensitive cancellation of large vector and scalar potentials amplifies and 
overemphasizes a 
(small) deviation from exact symmetry relations. 
Fortunately,  explicit symmetry breaking can be introduced in the nonlinear 
realization without affecting, e.g., the 
PCAC-relations. This allows for a parameterization of the hyperon potentials. 
Here, the term 
\be
\label{hyesb}
  {\cal L}_{hyp} = m_3 \Tr \left(\ovl{B} B + \ovl{B} [B, S]\right)\Tr (X-X_0)
\ee
with the same $S_{b}^a = -\frac{1}{3}[\sqrt{3} (\lambda_8)_{b}^a-\delta_{b}^a]$ as in Sec
\ref{bmver} is used.
The 
explicit symmetry breaking term  contributes 
only for hyperons at finite baryon densities along the hypercharge direction. 
With the parameter $m_3$ adjusted to the $\Lambda$ potential of $-28$ MeV, 
the other hyperon potentials are determined. This leads to 
a repulsive $\Xi$ potential ranging from $10-30$ MeV (table \ref{pots}).
We do not take the numbers for the $\Xi$ central potential too seriously because 
of the strongly varying values depending on the specific model and 
on the choice of the explicit symmetry breaking term. 
\section{Conclusions}
We studied a chiral SU(3) $\sigma$-$\omega$-type model including the 
dilaton associated 
with broken scale invariance of QCD.  Within such an approach it is 
possible to describe 
the multiplets of spin-0, spin-1, and spin-1/2 particles with reasonable
values for their vacuum masses as well as the nuclear matter equilibrium point 
at $\rho_0=0.15$ fm$^{-3}$ and the properties (e.g. binding energies, 
single particle energy spectra, charge radii) of nuclei. In contrast to 
other approaches to the nuclear many-body problem, all hadron masses 
are mainly generated through spontaneous symmetry breaking leading 
to a nonzero vacuum expectation value of a nonstrange ($\sigma$) and a 
strange ($\zeta$) condensate. 

In the linear $\sigma$ model, 
the vacuum expectation value of those two condensates is constrained 
by the decay constants of the pion ($f_\pi$) and of the kaon ($f_K$). 

It was shown, however that a SU(3) chiral model in the 
linear representation of chiral symmetry 
fails  to simultaneously account for nuclei and hyperon 
central potentials (see also \cite{paper2}). 
With that approach, it is either possible to 
describe nuclei with unrealistically low/high hyperon potentials 
{\it or} nuclear matter with reasonable hyperon potentials. This limitation 
does not exist if one switches to the nonlinear realization of chiral 
symmetry\footnote{However, we kept some constraints from 
the linear $\sigma$ model as e.g. the dependence of the condensates 
on the decay constants in order to reduce the amount of free parameter.}.
This has the following reasons:
\begin{itemize}
\item Firstly, an $f$-type baryon-scalar meson interaction 
can be constructed which does not destroy the balance between huge 
attractive and repulsive forces from the scalar and vector sector, 
respectively. This type of interaction improves the values for the hyperon 
potentials, though they remain too attractive. 
\item Secondly, the nonstrange and strange condensates can be decoupled 
 from each other, which reduces the level of attraction from 
the strange condensate. However, a decoupling of those condensates 
leads to masses for the whole scalar multiplet of the order of 500 MeV. 
A coupling of the condensates implying a mixing of the $\sigma$-and 
$\zeta$ scalar masses is necessary for a correct description of 
the hadronic spectrum. 
\item In contrast to the linear representation of chiral symmetry, 
it is possible to add an explicit symmetry breaking term which 
reduces the depth of the hyperon potentials without destroying basic 
theorems in the vacuum as the PCAC relations for the pseudoscalar 
mesons. However, in that direction further work has to be done 
to reduce the ambiguity of the explicit symmetry breaking term.
\end{itemize}

Within the nonlinear realization of chiral symmetry 
one also has the flexibility to construct some special potentials 
(in which the nonstrange/strange sector are decoupled from each other) 
and to reproduce the results of SU(2) models, as e.g. those 
obtained with the SU(2) model of the Minnesota-group \cite{heid94} and 
the nonlinear Boguta-Bodmer model \cite{boguta77}. 

However, to account for the scalar nonet masses, it is 
necessary to include terms which couple the nonstrange to the 
strange condensate. Particularly, it is  
possible to describe reasonably vacuum hadron masses, 
nuclear matter and nuclei within a single chiral SU(3) model in the 
nonlinear realization of chiral symmetry using the potential and some 
constraints of the linear $\sigma$ model.

The results are similar, whether the strange condensate 
is allowed to couple to the nucleon, or not. However, only 
in the first case it 
is possible to reproduce the experimentally known baryon masses without 
an additional explicit symmetry breaking term except for the one 
which can be associated with the current quark masses and which
produces finite masses for the pseudoscalar bosons. 
If the nucleon mass is entirely generated by 
the nonstrange $\sigma$ condensate, some additional explicit 
symmetry breaking is necessary to account for the correct baryon masses. 

To improve our results, 
a direct fit to spherical nuclei, as was done in \cite{rufa88} 
has to be performed. This is currently under investigation \cite{becki}. 
Further studies 
are under way to investigate the effect of spin-3/2 resonances in hot and 
dense matter, the meson-baryon scattering and the chiral dynamics in 
transport models within {\it one single model} \cite{stehof}.

\begin{acknowledgements}
The authors are grateful to L. Gerland, K. Sailer and the late J. Eisenberg 
for fruitful discussions. This work was funded in part by Deutsche 
Forschungsgemeinschaft (DFG), Gesellschaft f\"ur Schwerionenforschung (GSI) and 
Bundesministerium f\"ur Bildung und Forschung (BMBF). J. Schaffner-Bielich
is financially supported by the Alexander von Humboldt-Stiftung. 
\end{acknowledgements}
\appendix
\section{}
\label{append}
The various hadron matrices used are (suppressing the Lorentz indices)
\be
\label{smatrix}
X=\frac{1}{\sqrt{2}}\sigma^a \lambda_a=
\left( \begin{array}{ccc}
   (a_0^0  +\sigma)/\sqrt{2} & a_0^+ & \kappa^+\\   
    a_0^- & (-a_0^0+\sigma)/\sqrt{2} & \kappa^0 \\
   \kappa^- & \ovl{\kappa^0}& \zeta 
\end{array} \right)
\ee
\be
\label{vmatrix}
V=\frac{1}{\sqrt{2}}v^a \lambda_a=
\left( \begin{array}{ccc}
   (\rho_0^0  +\omega)/\sqrt{2} & \rho_0^+ & K^{\ast +}\\   
    \rho_0^- & (-\rho_0^0+\omega)/\sqrt{2} & K^{\ast 0} \\
   K^{\ast -} & \ovl{K^{\ast 0}}& \phi
\end{array} \right)
\ee
\be
\label{bmatrix}
B=\frac{1}{\sqrt{2}}b^a \lambda_a=
\left( \begin{array}{ccc}
   \frac{\Sigma^0}{\sqrt{2}} +\frac{\Lambda^0}{\sqrt{6}}& \Sigma^+ & p\\   
    \Sigma^- & -\frac{\Sigma^0}{\sqrt{2}} +\frac{\Lambda^0}{\sqrt{6}} & n \\
   \Xi^- & \Xi^0& -2 \frac{\Lambda^0}{\sqrt{6}}
\end{array} \right)
\ee
for the scalar ($X$), vector ($V$), baryon ($B$) 
and similarly for the axial vector meson fields.
A pseudoscalar chiral singlet
$Y=\sqrt{2/3} \eta_0  \, 1{\hspace{-2.5mm}1}$ can be added separately, since 
only an octet is allowed to enter the exponential \ref{piexp}.

The notation refers to the particles of the listed by the Particle Data Group
(PDG),\cite{pdg96}, though we are aware of the difficulties to directly 
identify the scalar mesons with the physical particles \cite{sche98}. However, 
note that there is increasing evidence which supports the existence of a low-mass, 
broad scalar resonance, the $\sigma(560)$-meson, as well as a light strange scalar 
meson, the $\kappa(900)$ (see \cite{black98} and references therein).


There is experimental indication for a nearly ideal 
mixing between the octet and singlet states. Hence, the nine vector mesons 
are summarized in a single matrix. 
The relevant fields in the SU(2) invariant vacuum, $v^0_{\mu}$ and 
$v^8_{\mu}$ (corresponding to $\lambda_0$ and $\lambda_8$, respectively), 
are assumed to have the ideal mixing angle, 
$\sin \theta_v =\frac{1}{\sqrt{3}}$. This  yields:
\bea
\label{holz3.36}
  \phi_{\mu}   &=& v_{\mu}^8 \cos \theta_v - v_{\mu}^0 \sin \theta_v =
  \frac{1}{\sqrt{3}} (\sqrt{2} v^0_{\mu}+v_8^{\mu})\\ \no
  \omega_{\mu} &=& v_{\mu}^8 \sin \theta_v + v_{\mu}^0 \cos \theta_v 
 =\frac{1}{\sqrt{3}} (v^0_{\mu}- \sqrt{2} v_8^{\mu}) .  
\eea
Similarly, for the scalar mesons
\bea
\sigma&=&\frac{1}{\sqrt{3}}(\sqrt{2}\sigma^0+\sigma^8) \\ 
\zeta &=&\frac{1}{\sqrt{3}}(\sigma^0-\sqrt{2} \sigma^8)
\eea
is used, where $\sigma^0$
and $\sigma^8$ belong to $\lambda_0$ and $\lambda_8$, respectively. 
However, there is no experimental indication for an ideal mixing of the  
scalar mesons $\sigma$ and $\zeta$. In general, depending 
on the interaction potential, mixing between $\sigma$ and $\zeta$ occurs
(see Sec. \ref{szint}). This is also suggested by effective instanton-induced
interactions of the 'tHooft type\cite{Hooft}. 

The masses of the various hadrons are generated through their couplings to 
the scalar condensates, which are produced via spontaneous symmetry
breaking in the sector of the scalar fields. 
There are nonvanishing vacuum expectation values (VEV) of 
only two meson fields: 
of the 9 scalar mesons in the matrix $X$ only the VEV of the 
components proportional to $\lambda_0$ and to the 
hypercharge $Y \sim \lambda_8$ are nonvanishing, and the vacuum expectation 
value $\langle X \rangle$ reduces to: 
\be
\label{vev}
\langle X \rangle=\frac {1}{\sqrt{2}}(\sigma^0 \lambda_0+\sigma^8 \lambda_8)
\equiv \mbox{diag } (\frac{\sigma}{\sqrt{2}} \,, \frac{\sigma}{\sqrt{2}} \,, 
\zeta ) , 
\ee
in order to preserve parity invariance and 
assuming, for simplicity, $SU(2)$ symmetry\footnote{This implies that 
isospin breaking effects will not occur, i.e., all hadrons of the 
same isospin multiplet will have identical masses. The electromagnetic 
mass breaking is neglected.} of the vacuum. 

\bibliography{chiral}
\bibliographystyle{prsty}
\clearpage
\newpage
\begin{table}
\begin{center}
\bt{c|c|c|c|c}
     & $m_{a_0} (980)$ & $m_{\kappa}$ (900)  & $m_{\sigma}$ & $m_{f_0}$  \\ \hline
C$_1$ &   953.54 &  995.70 &  473.32  & 1039.10 \\
C$_2$ &   953.54 &  995.70 &  475.55  & 1039.10  \\
C$_3$ &   953.54 &  995.70 &  478.56  & 824.17   \\
M$_1$ &   482.41 &  448.96 &  422.79  & 482.41  \\
M$_2$ &   488.55 &  441.41 &  408.79  & 488.55  \\
W$_1$ &   500.83 &  457.49 &  401.54 &  500.83 \\
W$_2$ &   519.95 &  478.34 &  425.14 &  519.95 \\
W$_3$ &   1000.00&  1255.32&   480.50 & 1334.25 \\
 \et
\caption{\label{massen} Vacuum masses of the scalar mesons for
different kinds of fits (explained in the text).}
\end{center}
\end{table}
\begin{table}
\begin{center}
\bt{c|c|c|c|c}
      & $m_N^\ast/m_N$ & $\sigma/\sigma_0$ & $\zeta/\zeta_0$ & $K$  \\ \hline
C$_1$ &   .61  &    .63   &   .92  & 276.34  \\
C$_2$ &   .64  &    .64   &   .91  & 266.08  \\
C$_3$ &   .61  &    .63   &   .92  & 285.29  \\
M$_1$ &   .62  &    .62   &  1   & 269.58  \\
M$_2$ &   .61  &    .62   &  1.01  & 272.61  \\
W$_1$ &   .65  &    .62   &  1   & 224.23        \\
W$_2$ &   .63  &    .63   &  1   & 245.05  \\
W$_3$ &   .64  &    .64   &   .91  & 217.20 \\
 \et
\caption{\label{nucmat} Condensates and nuclear matter properties at $\rho_0$.}
\end{center}
\end{table}

\begin{table}
\begin{center}
\bt{c|c|c|c|c|c}
      & U$_N$ & U$_{\Lambda}$ & U$_{\Sigma}$ & U$_{\Xi}$ & a$_4$\\ \hline

C$_1$ &-71.04   &-28.23   &  3.17   & 30.3    &40.41 \\
C$_2$ &  -68.75 &  -30.50 &   -6.46 & 21.1  &  37.29 \\
C$_3$ & -71.06  & -28.61  &   2.56  & 29.4   & 40.23 \\
M$_1$ &-70.18    & -46.78 & -46.78   &-23.39   & 40.59 \\
M$_2$ &-70.67    &-47.96  & -28.71   &-15.62   & 41.21 \\
W$_1$ &    -68.84  & -48.87 &  -42.92 &  -25.92  &  37.92     \\
W$_2$ &    -69.02  & -46.01 &  -46.01 &  -23.01  &  36.06     \\
W$_3$ &  -68.21    &-28.10  & -28.10  &  12.0  & 35.22  \\
 \et
\caption{\label{pots} Baryon potentials and asymmetry energy at
$\rho_0$. The hyperon potentials of the fits C$_1$, C$_2$, C$_3$, and 
W$_3$ are corrected with the explicit symmetry breaking term of Eq. \ref{hyesb}.}
\end{center}
\end{table}

\begin{table}
\begin{center}
\bt{c|ccc|ccc|ccc}
      & \multicolumn{3}{c}{$^{16}$O} & \multicolumn{3}{c}{$^{40}$Ca} &\multicolumn{3}{c}{$^{208}$Pb}\\ 
      &      E/A & r$_{ch}$ &  $\delta p$         & E/A & r$_{ch}$ & $\delta d$         & E/A & r$_{ch}$  & $\delta d$\\ \hline
Exp.  & -7.98    &  2.73    & 5.5-6.6 & -8.55  & 3.48 & 5.4-8.0 & -7.86  &  5.50 & 0.9-1.9  \\
      &          &          & &        &       &    &        &       &   \\
C$_1$ &  -7.30   &  2.65    & 6.05    &-7.98   &  3.42 &    6.19 &   -7.56 &    5.49  &   1.59 \\
C$_2$ &  -7.40   &  2.65    & 5.21    &-8.07   &  3.42 &    5.39 &   -7.61 &    5.50  &   1.41 \\
C$_3$ &  -7.29   &  2.65    & 6.06    &-7.98   &  3.42 &    6.22 &   -7.54 &    5.49  &   1.61 \\
M$_1$ &  -7.19   &  2.68    & 5.60    &-7.93   &  3.45 &    5.83 &   -7.56 &    5.53  &   1.53 \\
M$_2$ &  -7.34   &  2.67    & 5.90    &-8.03   &  3.44 &    6.08 &   -7.61 &    5.52  &   1.58 \\
W$_1$ &  -8.28   &  2.63    & 5.83    &-8.63   &  3.42 &    5.91 &   -7.71 &    5.51  &   1.43 \\
W$_2$ &  -8.23   &  2.63    & 5.84    &-8.60   &  3.42 &    5.94 &   -7.75 &    5.51  &   1.45 \\
W$_3$ &  -7.98   &  2.67    & 5.23    &-8.47   &  3.44 &    5.45    & -7.72 &    5.55 &    1.33 \\
 \et
\caption{\label{nuclei}
Bulk properties of nuclei:Prediction (left) and experimental values  
 (right) for
binding energy $E/A$, charge radius $r_{ch}$, and spin-orbit splitting
 of Oxigen ($^{16}$O with $\delta p\equiv p_{3/2}-p_{1/2}$), Calcium
($^{40}$Ca with $\delta d\equiv d_{5/2}-d_{3/2}$) and Lead ($^{208}$Pb with
$\delta d\equiv 2d_{5/2}-2d_{3/2}$).}
\end{center}
\end{table}

\begin{table}
\caption{\label{parameter} }
\bt{cccccccc}
    & k$_0$ & k$_1$ & k$_2$  & k$_3$ &k$_{3m}$ & k$_4$& $33 \delta$\\ \hline 
 C$_1$ & 2.37   &   1.40 &   -5.55  &  -2.65 &      0 &    -.23 &  2  \\
 C$_2$ & 2.36   &   1.40 &   -5.55  &  -2.64 &      0 &    -.23 &  2 \\
 C$_3$ & 2.35   &   1.40 &   -5.55  &  -2.60 &      0 &    -.23 &  2 \\
 M$_1$ & 1.28   &    0   &       0  &     0  &      0 &      0 &   6  \\ 
 M$_2$ & 1.29   &   0    &       0  &     0  &      0 &      0 &   6  \\
 W$_1$ & -14.91 &   0    &  16.67   &     0  &  32.06  &     0  &  0  \\ 
 W$_2$ & -12.96 &   0    &  16.67   &     0  &  32.06  &     0  &  0  \\
 W$_3$ & 10.44  &   7.32 &   -4.96  &     0  &   31.06 &     0 &   0 \\
   \et
\caption{\label{paramet} Parameters of the different potentials used (see text).}
\end{table}
\end{document}